\documentclass[twocolumn,10pt,letterpaper]{article}
\usepackage[top=2.3cm, bottom=2.3cm, left=1.8cm, right=1.8cm]{geometry}
\usepackage[compact]{titlesec}
\titlespacing*{\section}{0pt}{8pt}{4pt}
\titlespacing{\subsection}{0pt}{8pt}{4pt}
\titlespacing{\subsubsection}{0pt}{8pt}{2pt}

\usepackage{color,xcolor}
\usepackage{subfigure}
\usepackage{times}
\usepackage{cleveref}
\usepackage{balance}
\usepackage{graphicx}
\usepackage{url}
\usepackage{multirow}
\usepackage{graphicx}
\usepackage{amssymb}
\usepackage{amsfonts}
\usepackage[small,bf]{caption}
\usepackage{wrapfig}
\usepackage{alltt}
\usepackage{booktabs}
\usepackage{xspace}
\usepackage{verbatim}
\usepackage{footnote}
\usepackage{epstopdf}
\usepackage{rotating}
\definecolor{darkgreen}{RGB}{47,109,79}
\definecolor{darkblue}{RGB}{57,79,99}
\usepackage[bookmarks =false, colorlinks=true, plainpages = false,  linkcolor = darkgreen,   citecolor = darkgreen, urlcolor = darkblue, filecolor = blue]{hyperref}

\usepackage{flushend}
\newcommand{\etal}{\frenchspacing{}{et al{.}}\xspace}

\newcommand{\descr}[1]{\smallskip\noindent\textbf{#1}}

\usepackage[hang,flushmargin,stable]{footmisc}

\usepackage{indentfirst}

\usepackage{titling}
\makeatletter
\def\url@foostyle{%
  \@ifundefined{selectfont}{\def\UrlFont{\rm}}{\def\UrlFont{\rmfamily}}}
\makeatother
\makeatletter
\def\@copyrightspace{\relax}
\makeatother
\begin{document} 
\pagenumbering{arabic}

\title{\bf Measuring, Characterizing, and Detecting Facebook Like Farms\thanks{A preliminary version of this paper, titled ``Paying for Likes? Understanding Facebook Like Fraud Using Honeypots''~\cite{decristofaro14facebooklikefarms}, appeared in ACM Internet Measurement Conference 2014 (IMC'14). Please see the last paragraph in Section~\ref{sec:related} (page 17) for a summary of the new results presented in this version, which is published in ACM Transactions on Privacy and Security (TOPS).}}

\author{Muhammad Ikram$^{1\dag}$, Lucky Onwuzurike$^{2\dag}$, Shehroze Farooqi$^3$,
Emiliano De Cristofaro$^2$,\\ Arik Friedman$^{4\ddag}$, Guillaume Jourjon$^1$,
Mohamed Ali Kaafar$^1$, M. Zubair Shafiq$^3$\\[0.75ex]
\normalsize $^1$Data61-CSIRO $\;\;\;$ $^2$University College London $\;\;\;$ $^3$University of Iowa $\;\;\;$ $^4$Atlassian \vspace*{-1cm}}

\date{}
\maketitle

\renewcommand{\thefootnote}{\fnsymbol{footnote}}
\footnotetext{$^\dag$Authors contributed equally.\\$^\ddag$Work done while the author was with Data61-CSIRO.}
\renewcommand{\thefootnote}{\arabic{footnote}}

\begin{abstract}
Online social networks offer convenient ways to seamlessly reach out to large audiences. In particular, Facebook pages are increasingly used by businesses, brands, and organizations to connect with multitudes of users worldwide. As the number of likes of a page has become a de-facto measure of its popularity and profitability, an underground market of services artificially inflating page likes, aka {\em like farms}, has emerged alongside Facebook's official targeted advertising platform. Nonetheless, besides a few media reports, there is little work that systematically analyzes Facebook pages' promotion methods. 
Aiming to fill this gap, we present a honeypot-based comparative measurement study of page likes garnered via Facebook advertising and from popular like farms. First, we analyze likes based on demographic, temporal, and social characteristics, and find that some farms seem to be operated by bots and do not really try to hide the nature of their operations, while others follow a stealthier approach, mimicking regular users' behavior. Next, we look at fraud detection algorithms currently deployed by Facebook and show that they do not work well to detect stealthy farms which spread likes over longer timespans and like popular pages to mimic regular users. 

To overcome their limitations, we investigate the feasibility of timeline-based detection of like farm accounts, focusing on characterizing content generated by Facebook accounts on their timelines as an indicator of genuine versus fake social activity. We analyze a wide range of features extracted from timeline posts, which we group into two main categories: lexical and non-lexical. We find that like farm accounts tend to re-share content more often, use fewer words and poorer vocabulary, and more often generate duplicate comments and likes compared to normal users. Using relevant lexical and non-lexical features, we build a classifier to detect like farms accounts that achieves a precision higher than 99\% and a 93\% recall.
\end{abstract}

\section{Introduction}\label{sec:introduction}

Online social networks provide organizations and public figures with a range of tools to reach out to/broaden their audience. Among these, {\em Facebook pages} make it easy to broadcast updates, publicize products and events, and get in touch with customers and fans. Facebook allows page owners to promote their pages via targeted advertisement, i.e., pages can be ``suggested'' to  users from  specific age or location groups, or with certain interests. Page ads constitute one of the primary sources of revenue for Facebook, as its advertising platform overall is reportedly used by 2 million small businesses, out of the 40 million which have active pages~\cite{facebookpagecount}.

At the same time, as the number of likes on a Facebook page is considered a measure of its popularity~\cite{carter13like}, an ecosystem of so-called {\em ``like farms''} has emerged that offers paid services to artificially inflate the number of likes on Facebook pages. These farms rely on fake and compromised accounts as well as incentivized collusion networks where users are paid for actions from their account~\cite{viswanath14tanomaloussocialnetwork}.
Popular media reports~\cite{bbc,dangerousminds,guardianreport,veritasiumfbfraud,likeorlie} have speculated that Facebook ad campaigns may also garner significant amounts of fake likes, due to farm accounts' attempt to diversify liking activities and avoid Facebook's fraud detection algorithms.
With the price charged by like farms varying, for 1000 likes, from \$14.99--\$70 for worldwide users to \$59.95--\$190 for USA users, it is not far fetched to assume that selling likes may yield significant profits for fraudsters.
This also creates potential problems for providers like Facebook as they lose potential ad revenues while possibly disenfranchising page owners who receive likes from users who do not engage with their page.
However, even though the understanding of fake likes is crucial to improve fraud mitigation in social networks, there has been little work to systematically analyze and compare Facebook page promotion methods.
With this motivation in mind, we set to shed light on the like farming ecosystem with the aim of characterizing features and behaviors that can be useful to effectively detect them. In the process, we review the fraud detection tools currently deployed by Facebook and assess their efficacy for more sophisticated like farms.

Specifically, our paper makes three main contributions:
\begin{enumerate}
\item We present a first-of-its-kind honeypot-based comparative measurement study of page likes garnered via Facebook ads and like farms, and  analyze likes based on demographic, temporal, and social characteristics;
\item We perform an empirical analysis of graph-based fraud detection tools used by Facebook and highlight their shortcomings against more sophisticated farms; and
\item We propose and evaluate timeline-based detection of like farm accounts, focusing on characterizing content as an indicator of genuine versus fake social activity, and build a classifier, based on lexical and non-lexical features, that detects like farm accounts with at least 99\% precision and 93\% recall.
\end{enumerate}

\subsection{Roadmap}
\descr{Honeypot-based measurement of like farms.} Aiming to study fake likes garnered from like farms and, potentially, from Facebook advertising, we create thirteen Facebook {\em honeypot} pages with the description: ``This is not a real page, so please do not like it.'' and intentionally kept them empty  (i.e.,  no posts or pictures). We promote eight using four like farms (i.e., targeting users in the USA and worldwide for each, as farms mostly offer user targeting for only this two locations) and five using Facebook ad campaigns (with two targeting users in the USA and worldwide as the like farms. The other three target one developed and two developing countries, as Facebook reports that ``false'' accounts are less prevalent in developed markets and
more in developing markets.\footnote{\url{https://goo.gl/OAxgTh}.}
After monitoring likes garnered by the pages, and collecting information about the likers
(e.g., gender, age, location, friend list, etc.), we perform a comparative analysis based on demographic, temporal, and social characteristics.

We identify two main {\em modi operandi} for the like farms: (1) some seem to be  operated by bots and do not really try to hide their activities, delivering likes in bursts and forming disconnected social sub-graphs, while (2) others follow a stealthier approach, mimicking regular users' behavior, and rely on a large and well-connected network structure to gradually deliver likes while keeping a small count of likes per user. The first strategy reflects a ``quick and dirty'' approach where likes from fake
users are delivered rapidly, as opposed to the second one,  which exhibits a stealthier approach that leverages the underlying social graph, where accounts (possibly operated by real users) slowly deliver likes.
We also highlight a few more interesting findings. When targeting Facebook users worldwide, we obtain likes from only a few countries, and that likers' profiles seem skewed toward males. Moreover, we find evidence that different like farms (with different pricing schemes) garner likes from overlapping sets of users and, thus, may be managed by the same operator.

\descr{Characterizing fake likes.} We present the concept of liking a page on Facebook as a binary action where likes received on a page by users who have interest for the content of the page are considered ``good'' and likes received in order to manipulate a page's popularity ranking as ``fake''. We have only considered to mark as fake, likes that are meant to manipulate a page's popularity (i.e., by increasing the page's number of fans) as this is the main purpose like farms serve. On this note, we start our study with the assumption that likers from farms that like our empty honeypot pages are either fake or compromised real users (i.e., fake likes) as shown in~\cite{viswanath14tanomaloussocialnetwork}. Although, Facebook discourages page owners from buying fake likes, warning that they {\em ``can be harmful to your page''}\footnote{See \url{https://www.facebook.com/help/241847306001585}.}, they also routinely launch clean-up campaigns to remove fake accounts, including those engaged in like farms. Hence, we also hypothesize that very few or no users from the Facebook Ad campaigns will like our honeypot pages as the pages were empty.

Aiming to counter like farms, researchers as well as Facebook have recently been working on tools to detect fake likes. One currently deployed tool is CopyCatch, which detects lockstep page like patterns by analyzing the social graph between users and pages, and the times at which the edges in the graph are created~\cite{beutel2013copycatch}. Another one, SynchroTrap, relies on the fact that malicious accounts usually perform loosely synchronized actions in a variety of social network context, and can cluster malicious accounts that act similarly at around the same time for a sustained period of time~\cite{cao14synchrotrap}. The issue with these methods, however, is that stealthier (and more expensive) like farms can successfully circumvent them by spreading likes over longer timespans and liking popular pages to mimic normal users.
We systematically evaluate the effectiveness of these graph-based co-clustering fraud detection algorithms~\cite{beutel2013copycatch,cao14synchrotrap} in identifying like farm accounts. We show that these tools incur  high false positives rates for stealthy farms, as their accounts mimic normal users.

\descr{Characterizing lexical and non-lexical timeline information.} Next, we investigate the use of timeline information, including lexical and non-lexical characteristics of user posts, to improve the detection of like farm accounts. To this end, we crawl and analyze timelines of user accounts associated with like farms as well as a baseline of normal user accounts. Our analysis of timeline information highlights several differences in both lexical and non-lexical features of baseline and like farm users. In particular, we find that timeline posts by like farm accounts have 43\% fewer words, a more limited vocabulary, and lower readability than normal users' posts. Moreover, like farm accounts' posts generate significantly more comments and likes, and a much larger fraction of their posts consists of ``shared activity'' (i.e., sharing posts from other users, news articles, videos, and external URLs).

\descr{Detection.} Based on our characterization, we extract a set of timeline-based features and use them to train three classifiers using supervised two-class support vector machines (SVM)~\cite{Muller01anintroduction}.  Our first and second classifiers use, respectively, lexical and non-lexical features extracted from timeline posts, while the third one uses both. We evaluate the classifiers using the ground-truth dataset of like farm accounts and show that they achieve 99--100\% precision and 93--97\% recall in detecting like farm accounts. Finally, we generalize our approach using other classification algorithms, namely, decision tree~\cite{dtree}, AdaBoost~\cite{adaboost}, kNN~\cite{knn}, random forest~\cite{Breiman:rf}, and na\"ive Bayes~\cite{zhang2004optimality}, and empirically confirm that the SVM classifier achieves higher accuracy across the board.

\subsection{Paper organization}
 The rest of the paper is organized as follows. Section~\ref{sec:imc} presents our honeypot-based comparative measurement of likes garnered using farms and legitimate Facebook ad campaigns. Then, Section~\ref{sec:coclustering} evaluates the accuracy of state-of-the-art co-clustering techniques to detect like farm accounts in our datasets. Next, we study timeline based features (both non-lexical and lexical) in Section~\ref{sec:characterizing}, and evaluate the classifiers built using these features in Section~\ref{sec:detection}. After reviewing related work in Section~\ref{sec:related}, the paper concludes in Section~\ref{sec:conclusion}.

\begin{table*}[t!]
\small
\centering
		  \resizebox{0.8\textwidth}{!}{%
\begin{tabular}{lllrrrrr}
\toprule
  \bf{Campaign} & \bf{Provider} & \bf{Location} & \bf{Budget} & \bf{Duration} & {\bf Moni-} &
  \bf{\#Likes} & {\bf \#Termi-}\\
  {\bf ID} & & & & & {\bf toring} &  & {\bf nated}\\
\midrule
FB-USA & Facebook.com  & USA & \$6/day  & 15 days & 22 days & 32 & 0\\
FB-FRA & Facebook.com &  France & \$6/day & 15 days &22 days & 44 & 0 \\
FB-IND & Facebook.com &  India & \$6/day  & 15 days &22 days & 518 & 2 \\
FB-EGY & Facebook.com &  Egypt & \$6/day & 15 days &22 days & 691 & 6\\
FB-ALL & Facebook.com &  Worldwide & \$6/day  & 15 days &22 days &484 & 3\\
\midrule
BL-ALL & BoostLikes.com    & Worldwide & \$70.00 & 15 days &-& - & -\\
BL-USA & BoostLikes.com   & USA only & \$190.00 & 15 days &22 days & 621 & 1\\
SF-ALL & SocialFormula.com   & Worldwide & \$14.99 & 3 days  &10 days & 984 & 11\\
SF-USA & SocialFormula.com  & USA & \$69.99 & 3 days &10 days & 738 & 9\\
AL-ALL & AuthenticLikes.com    & Worldwide & \$49.95 & 3-5 days & 12 days &755 & 8 \\
AL-USA & AuthenticLikes.com   & USA & \$59.95& 3-5 days &22 days & 1038 & 36 \\
MS-ALL & MammothSocials.com  & Worldwide & \$20.00 & -  &-  & - & -\\
MS-USA & MammothSocials.com  & USA only & \$95.00 & - &12 days & 317 & 9\\
\bottomrule
\end{tabular}
}
\vspace{-0.15cm}
\caption{Facebook and like farm campaigns used to promote the Facebook honeypot pages. Like farms promised to deliver 1000 likes in 15 days at differing prices depending on the geographical target (i.e., USA and worldwide) whereas on Facebook, we budgeted \$6 per day for the promotion of each page for a period of 15 days.}
\label{tbl:measurements-imc}
\end{table*}

\section{Honeypot-based Measurement of Facebook Like Farms}
\label{sec:imc}
This section details our honeypot-based comparative measurement study of page likes garnered via Facebook ads and by like farms.

\subsection{Datasets}\label{sec:data-imc}
In the following, we present the methodology used to deploy, monitor, and promote our Facebook honeypot pages.%

\descr{Honeypot Pages.} In March 2014, we created 13 Facebook pages called ``Virtual Electricity'' and intentionally kept them empty (i.e., no posts or pictures). Their description included: {\em ``This is not a real page, so please do not like it.''}
5 pages were promoted using legitimate Facebook (FB) ad campaigns targeting users, respectively, in USA, France, India, Egypt, and worldwide.
The remaining 8 pages were promoted using 4 popular like farms: BoostLikes.com (BL), SocialFormula.com (SF), AuthenticLikes.com (AL), and MammothSocials.com (MS), targeting worldwide or USA users.

In Table \ref{tbl:measurements-imc}, we provide details of the honeypot pages, along with the
corresponding ad campaigns. All campaigns were launched on March 12, 2014, using a different administrator account (owner) for each page. Each Facebook campaign was budgeted at a maximum of \$6/day to a total of \$90 for 15 days. The price for buying likes varied across like farms: BoostLikes charged the highest price for ``100\% real likes'' (\$70 and \$190 for 1000 likes in 15 days from, respectively, worldwide and USA).
Other like farms also claimed to deliver likes from ``genuine'', ``real'', and ``active'' profiles, but promised to deliver them in fewer days.
Overall, the price of 1000 likes varied between \$14.99--\$70 for worldwide users and \$59.95--\$190 for USA users.

\descr{Data Collection.} We monitored the ``liking'' activity on the honeypot pages by crawling them every 2 hours using Selenium web driver.
At the end of the campaigns, we reduced the frequency of monitoring to once a day, and stopped monitoring when a page did not receive a like for more than a week.
We used Facebook's reports tool for page administrators, which provides a variety of aggregated statistics
about attributes and profiles of page likers. Facebook also provides these statistics for the global Facebook population. Since a majority of Facebook users do not set the visibility of their age and location to public~\cite{Chaabane2012}, we used these reports to collect statistics about likers' gender, age, country, home and current town.
Later in this section, we will use these statistics to compare distributions of our honeypot pages' likers to that of the overall Facebook population.
We also crawled public information from the likers' profiles, obtaining the lists of liked pages as well as friend lists, which are not provided in the reports. Overall, we identify more than 6.3 million total likes by users who liked our honeypot pages and more than 1 million friendship relations. %

\descr{Campaign Summary.} In Table~\ref{tbl:measurements-imc}, we report the total number of likes garnered by each campaign, along with the number of days we monitored the honeypot pages. Note that the BL-ALL and MS-ALL campaigns remained inactive, i.e., they did not result in any likes even though we were charged in advance.
We tried to reach the like farm admins several times but received no response. Overall, we collected a total of 6,222 likes (4,453 from like farms and 1,769 from Facebook ads). The largest number of likes were garnered by AL-USA, the lowest (excluding inactive campaigns) by FB-USA.

\descr{Ethics Considerations.} Although we only collected openly available data,
we did collect (public) profile information from our honeypot pages' likers, e.g., friend lists and page likes. We could not request consent but enforced a few mechanisms to protect user privacy: all data were encrypted at rest and not re-distributed, and no personal information was extracted, i.e., we only analyzed aggregated statistics.
We are also aware that paying farms to generate fake likes might raise ethical concerns, however, this was crucial to create the honeypots and observe the like farms' behavior. We believe that the study will help, in turn, to understand and counter these activities. Also note that the amount of money each farm received was small (\$190 at most) and that this research was reviewed and approved by Data61's legal team. We also received ethical approval from the ethics committee of UCL where, in conjunction with Data61, data was collected and analyzed.

\subsection{Location and Demographics Analysis}
\label{sec:demographics}
We now set to compare the characteristics of the likes garnered by the honeypot pages promoted via legitimate Facebook campaigns and those obtained via like farms.

\begin{figure}[t!]
\centering
\includegraphics[width=0.9\columnwidth]{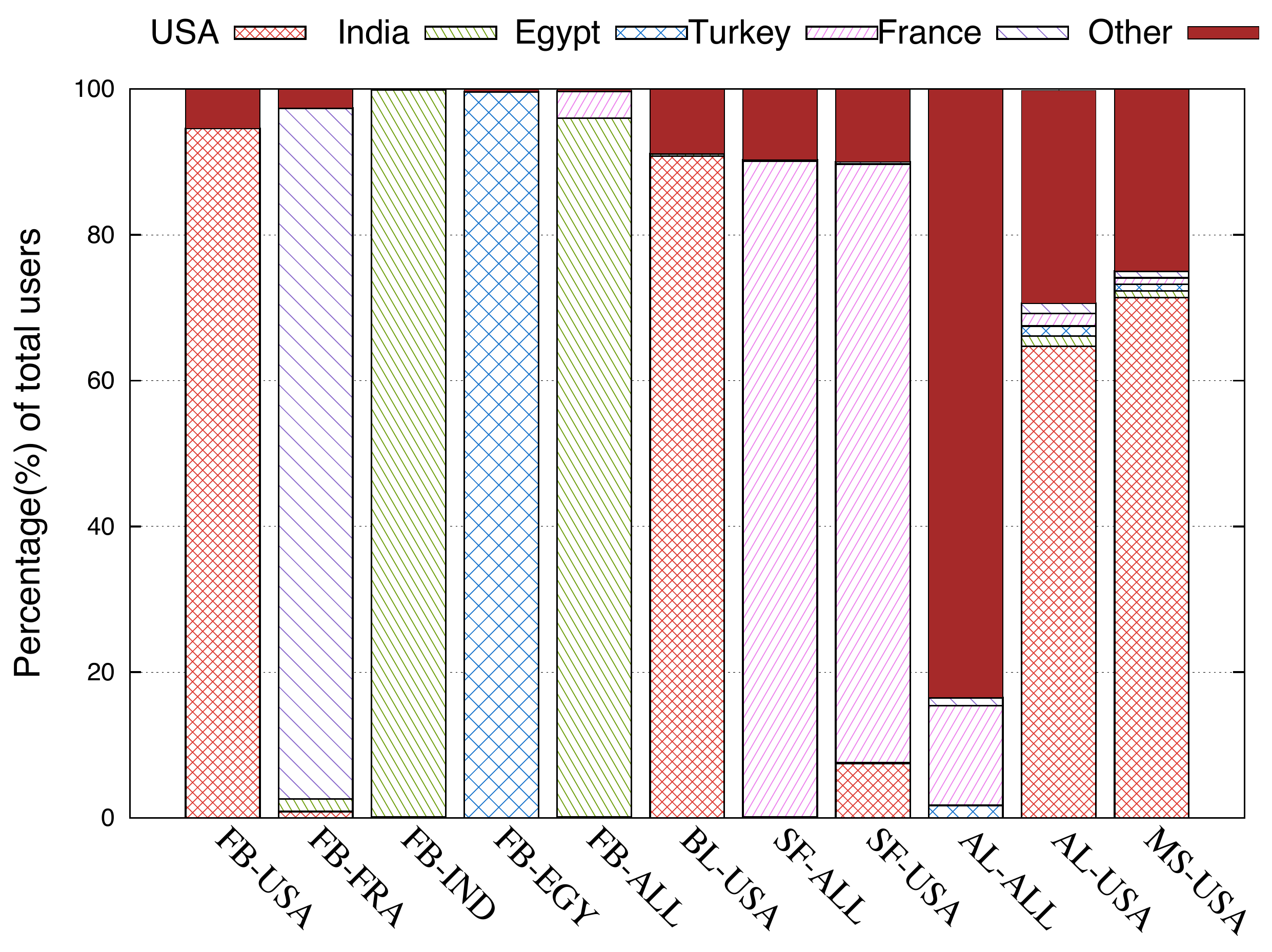}
\vspace{-0.15cm}
\caption{Geolocation of the likers (per campaign). }
\label{fig:geolocation}
\end{figure}

\descr{Location.} For each campaign, we looked at the distribution of likers' countries: as shown in Figure~\ref{fig:geolocation}, for the first four Facebook campaigns (FB-USA, FB-FRA, FB-IND, FB-EGY), we mainly received likes from the targeted country (87--99.8\%), even though FB-USA and FB-FRA generated a number of likes much smaller than any other campaign.
When we targeted Facebook users worldwide (FB-ALL), we almost exclusively received likes from India (96\%).
Looking at the like farms, most likers from SocialFormula were based in Turkey, regardless of whether we requested a US-only campaign. The other three farms delivered likes complying to our requests, e.g., for US-only campaigns, the pages received a majority of likes from US profiles. The location result supports Facebook's claim that ``the percentage of accounts that are duplicate or false is meaningfully lower in developed markets such as the United States or United Kingdom and higher in developing markets such as India and Turkey.''\footnote{\url{https://goo.gl/OAxgTh}.} It also potentially supports the claim that like farm accounts diversify their liking activities by liking pages promoted via Facebook ads to avoid Facebook's fraud detection algorithms (we further explore this in section~\ref{sec:likeAnalysis}).

\descr{Other Demographics.}
In Table~\ref{tbl:attributes}, we show the distribution of likers' gender and age, and also compare them to the global Facebook network (last row). The last column reports the KL-divergence between the age distribution of the campaign users and that of the entire Facebook population, highlighting large divergence for FB-IND, FB-EGY, and FB-ALL, which are biased toward
younger users. These three campaigns also appear to be skewed toward male profiles.
In contrast, the demographics of likers from SocialFormula and, to a lesser extent, AuhtenticLikes and MammothSocials, are much more similar to those of the entire network, even though male users are still over-represented. %

\begin{table}[t!]
\tabcolsep=0.11cm
\centering
		\resizebox{0.5\textwidth}{!}{%
\begin{tabular}{lcccccccc}
\toprule
 {\bf{Campaign}} &  {\bf Gender } & \multicolumn{6}{c}{\bf{-- Age Distribution (\%) --}} & \\
 {\bf{ID}} & {\bf \% F/M} &{\bf 13-17} & {\bf 18-24}  & {\bf 25-34} & {\bf 35-44} & {\bf 45-54}  & \multicolumn{1}{c}{\bf 55+} & {\bf KL}  \\
\midrule
 FB-USA& 54/46 & 54.0&	27.0&	6.8&	6.8&	1.4&	4.1 & 0.45 \\
 FB-FR	&46/54&60.8&	20.8&	8.7&	2.6&	5.2&	1.7 & 0.54 \\
FB-IND	&\bf{7/93}&52.7&	43.5&	2.3&	0.7&	0.5&	0.3 & \bf{1.12} \\
FB-EGY	&\bf{18/82}&54.6&	34.4&	6.4&	2.9&	0.8&	0.8 & \bf{0.64} \\
FB-ALL	&\bf{6/94}&51.3&	44.4&	2.1&	1.1&	0.5&	0.6 & \bf{1.04}\\
\midrule
BL-USA	&53/47&34.2&	54.5&	8.8&	1.5&	0.7&	0.5 & 0.60\\
SF-ALL	&37/63&19.8&	33.3&	21.0&	15.2&	7.2&	2.8 & 0.04\\
SF-USA	&37/63&22.3	&34.6&22.9&	11.6&	5.4&	2.9 & 0.04 \\
AL-ALL	&42/58&15.8&	52.8&	13.4&	9.7&	5.2&	3.0 & 0.12 \\
AL-USA	&31/68&7.2&	41.0&	35.0&	10.0&	3.5&	2.8 & 0.09\\
MS-USA	&26/74&8.6&	46.9&	34.5&	6.4&	1.9&	1.4 & 0.17\\
\midrule
Facebook	&46/54&14.9&	32.3&	26.6&	13.2&	7.2&	5.9 & -- \\
\bottomrule
\end{tabular}
}
\vspace{-0.15cm}
\caption{Gender and age statistics of likers.}%
\label{tbl:attributes}
\end{table}

\begin{figure*}[t!]
\centering
         \subfigure[\small Like Farms]{\includegraphics[width=0.74\columnwidth]{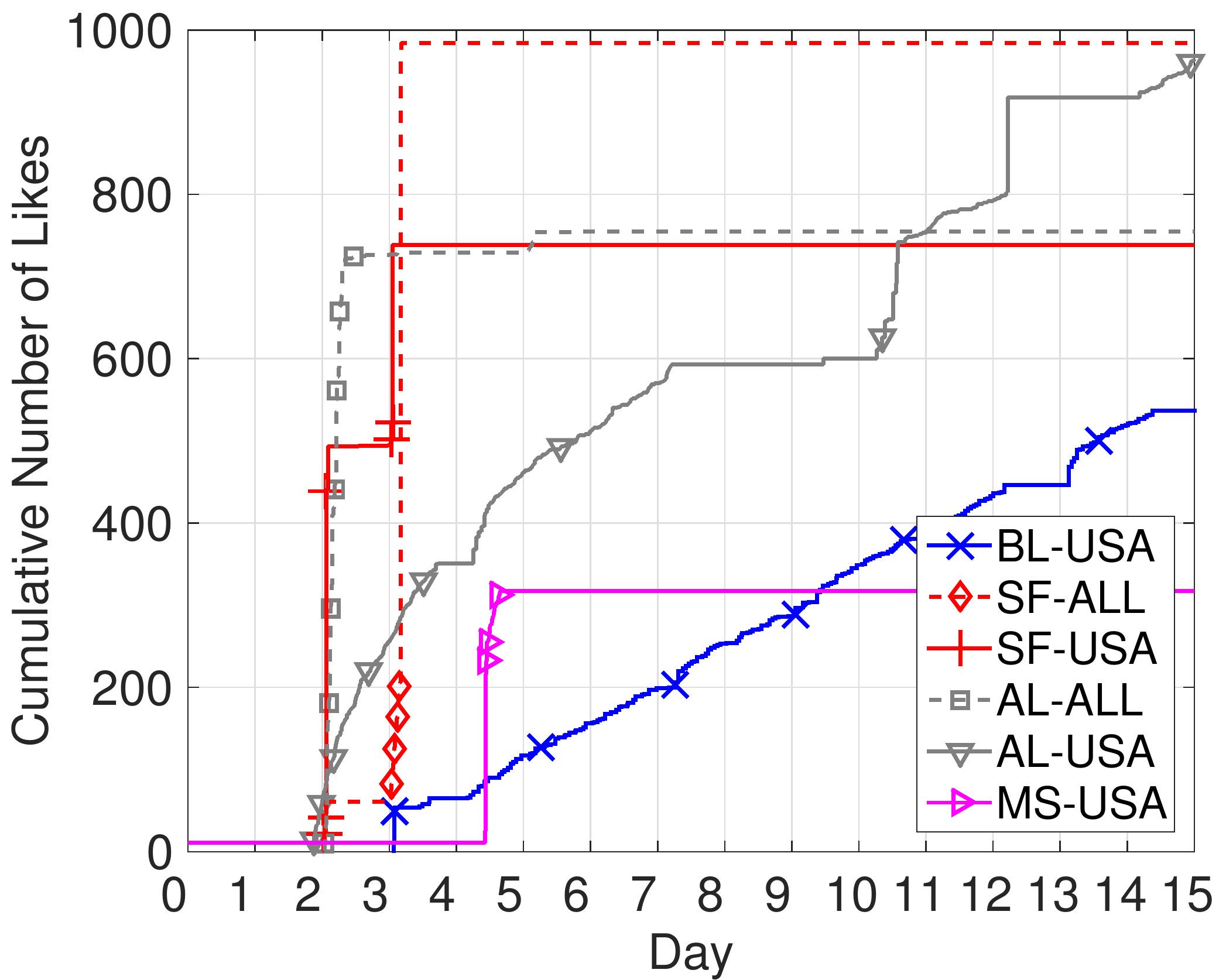}
         \label{likes_timeseries_underground}}
	\subfigure[\small Facebook Campaigns]{\includegraphics[width=0.74\columnwidth]{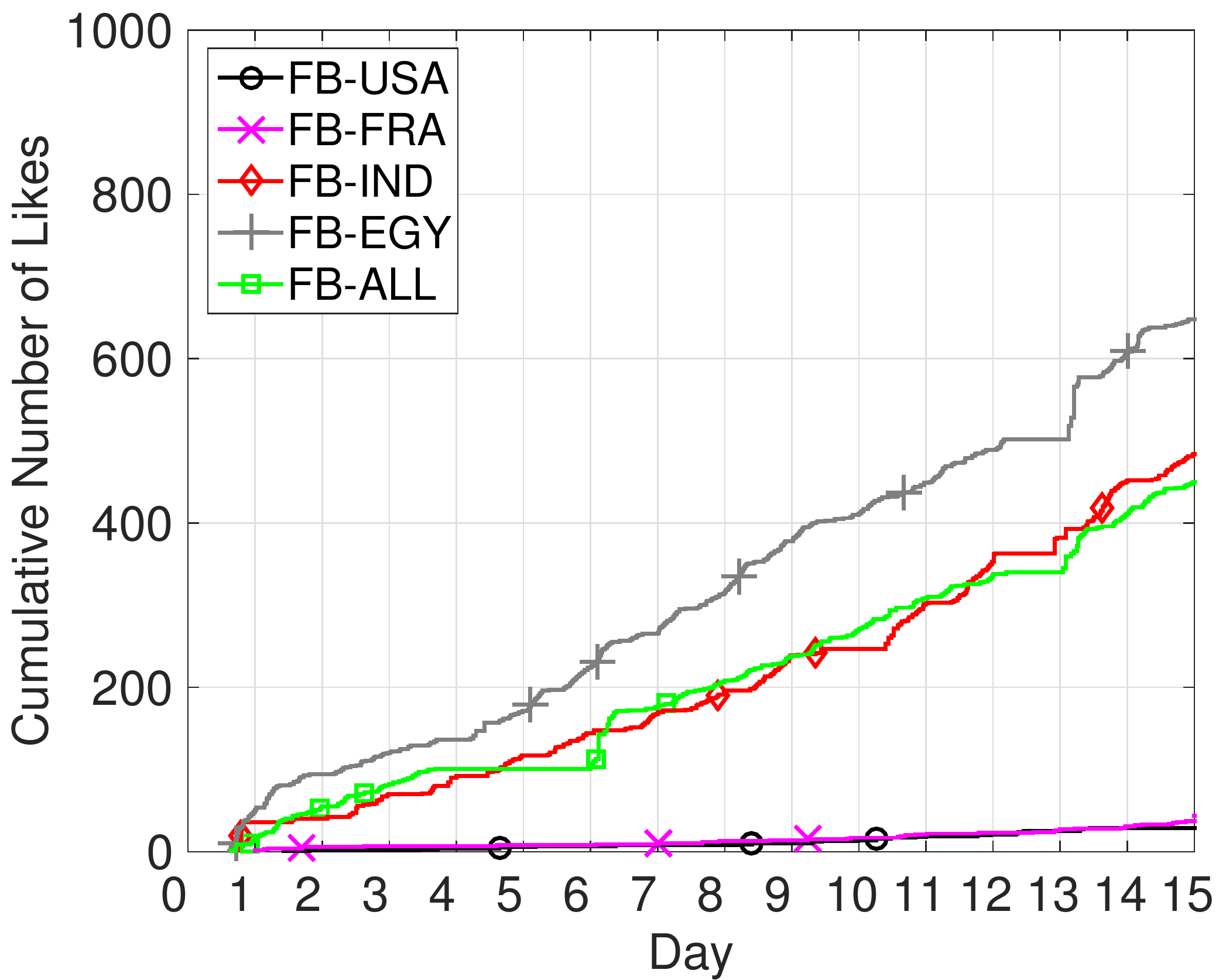}
         \label{likes_timeseries_facebook}}
\vspace{-0.15cm}
\caption{Time series of cumulative number of likes for Facebook and like farms campaigns.} %
\label{Fig:results:temporalAccumulcation}
\end{figure*}

\subsection{Temporal Analysis}
We also analyzed temporal patterns observed for each of the campaigns. In Figure~\ref{Fig:results:temporalAccumulcation}, we plot the cumulative number of likes observed on each honeypot page over our observation period (15 days). We observe from Figure~\ref{likes_timeseries_underground} that all the like farm campaigns, except BoostLikes, exhibit a very similar trend with a few bursts of a large number of likes. Specifically, for the SocialFormula, AuthenticLikes, and MammothSocials campaigns, likes were garnered within a short period of time of two hours. With AuthenticLikes, we observed likes from more than 700 profiles within the first 4 hours of the second day of data collection. Interestingly, no more likes were observed later on.
On the contrary, the BoostLikes campaign targeting US users shows a different temporal behavior: the trend is actually comparable to that observed in the Facebook Ads campaigns (see Figure~\ref{likes_timeseries_facebook}). The number of likes steadily increases during the observation period and no abrupt changes are observed.

This suggests that two different strategies may be adopted by like farms. On the one hand, the abrupt increase in the cumulative number of likes happening during a short period of time might likely be due to %
automated scripts operating a set of fake profiles. These profiles are instrumented to satisfy the number of likes as per the customer's request.
On the other hand, BoostLikes's strategy, which resembles the temporal evolution in Facebook campaigns,  seems to rely on the underlying social graph, possibly constituted by fake profiles operated by humans. Results presented in the next section corroborate the existence of these two strategies.

\subsection{Social Graph Analysis}

\begin{table*}[t!]
\tabcolsep=0.11cm
\small
\centering
		  \resizebox{0.7\textwidth}{!}{%
\begin{tabular}{lrrrrrr}
\toprule
\bf{Provider} & \bf{\#Likers} & \bf{\#Likers with} & \bf{Avg ($\pm$ Std)} & {\bf Median} & \bf{\#Friendships}  & \bf{\#2-Hop Friend-}  \\
 &  & \bf{Public Friend}  &  \bf{\#Friends} & {\bf \#Friends} & \bf{Between} & \bf{ship Relations}\\
 & & {\bf Lists} &  & & {\bf Likers} & {\bf Between Likers}\\
 \midrule
FB & 1448 &  261 (18.0\%) &  315 $\pm$ 454 & 198 &  6 & 169\\
BL & 621 &  161 (25.9\%) & 1171 $\pm$ 1096 &  850 & 540 & 2987  \\
SF & 1644 & 954 (58.0\%) &  246 $\pm$ 330& 155 &50 & 1132  \\
AL & 1597 &  680 (42.6\%) & 719 $\pm$ 973& 343 & 64 & 1174  \\
MS & 121 & 62 (51.2\%) & 250 $\pm$ 585& 68 & 4 & 129 \\
ALMS & 213 & 101 (47.4\%) &  426 $\pm$ 961 & 46 &  27 & 229  \\
\hline
\end{tabular}
}
\caption{Likers and friendships between likers.}
\label{tab:LikersAndFriendships}
\end{table*}

Next, we evaluated the social graph induced by the likers' profiles.
To this end, we associated each user with one of the like farm services based on the page they liked. Note that a
few users liked pages in multiple campaigns, as we will discuss in Section~\ref{sec:likeAnalysis}.
A significant fraction of users actually liked pages corresponding to both the AuthenticLikes and the MammothSocials campaigns (see Figure~\ref{fig:like-distribution}): we put these users into a separate group, labelled as ALMS. Table~\ref{tab:LikersAndFriendships} summarizes the number of likers associated with each service, as well as additional details about their friendship networks. Note that the number of likers reported for each campaign in Table~\ref{tab:LikersAndFriendships} is different from the number of campaign likes (Table~\ref{tbl:measurements-imc}), since some users liked more than one page.

Many likers kept their friend lists private: this occurred for almost 80\% of likers in the Facebook campaigns, about 75\% in the BoostLikes campaign, and much less frequently for the other like farm campaigns ($\sim$40--60\%). The number and percentage of users  with public friend lists are reported in Table~\ref{tab:LikersAndFriendships}.
The fourth column reports the average number of friends ($\pm$ the standard deviation) for profiles with visible friend lists, and the fifth column reports the median.
Some friendship relations may be hidden, e.g., if a friend chose to be invisible in friend lists, thus, these numbers only represent a {\em lower bound}. The average number of friends of users associated with the BoostLikes campaign (and to a smaller extent, the AuthenticLikes campaign) was much higher than the average number of friends observed elsewhere.

\begin{figure*}[t!]
\centering
         \subfigure[Direct friendship relations\label{fig:friendship}]{\includegraphics[width=0.65\columnwidth]{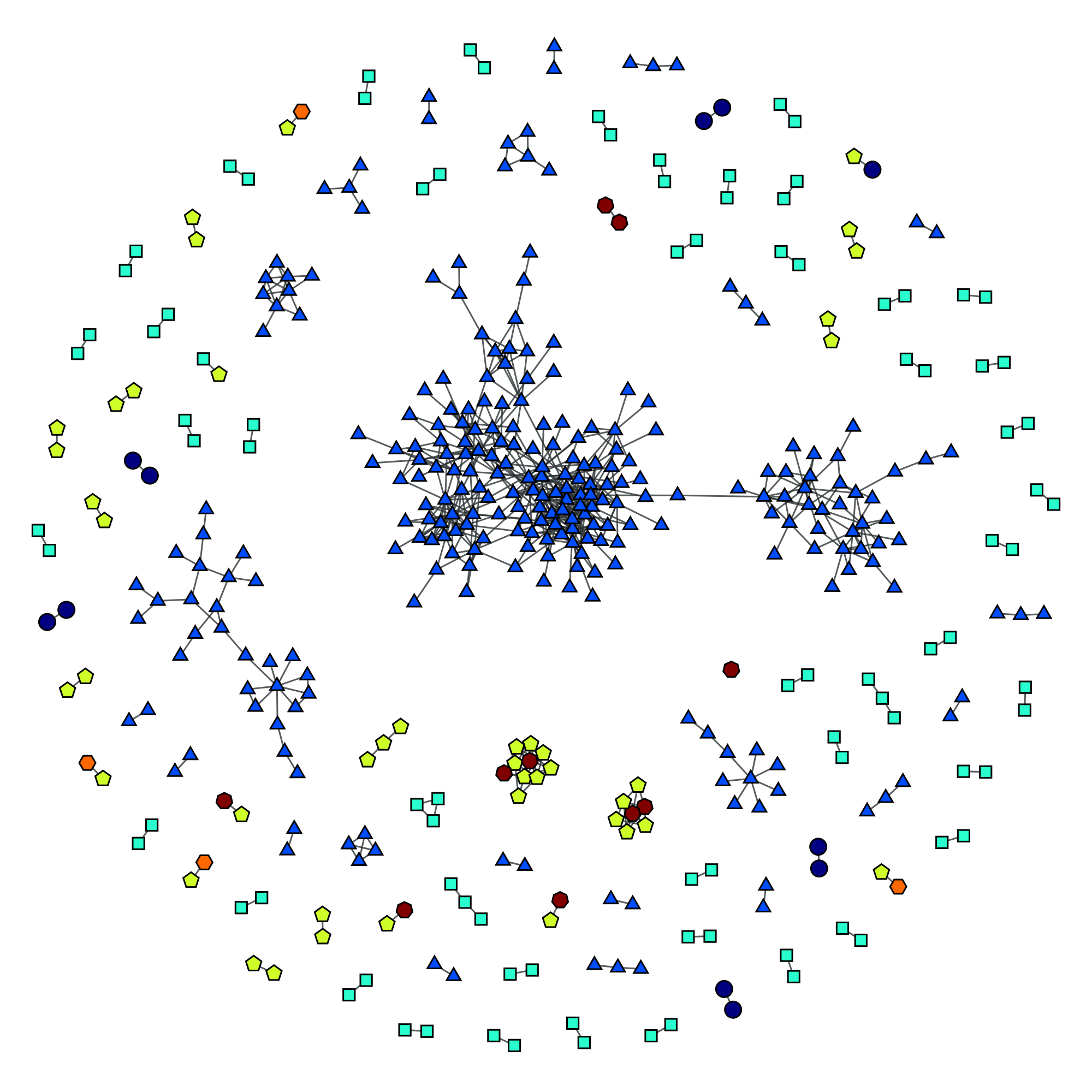}}\hspace*{-0.35cm}
         \subfigure{\includegraphics[width=0.35\columnwidth]{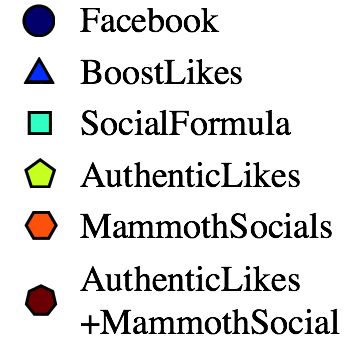}}
         \addtocounter{subfigure}{-1}\hspace*{-0.5cm}
         \subfigure[2-hop friendship relations\label{fig:2hop_friendship}]{\includegraphics[width=0.65\columnwidth]{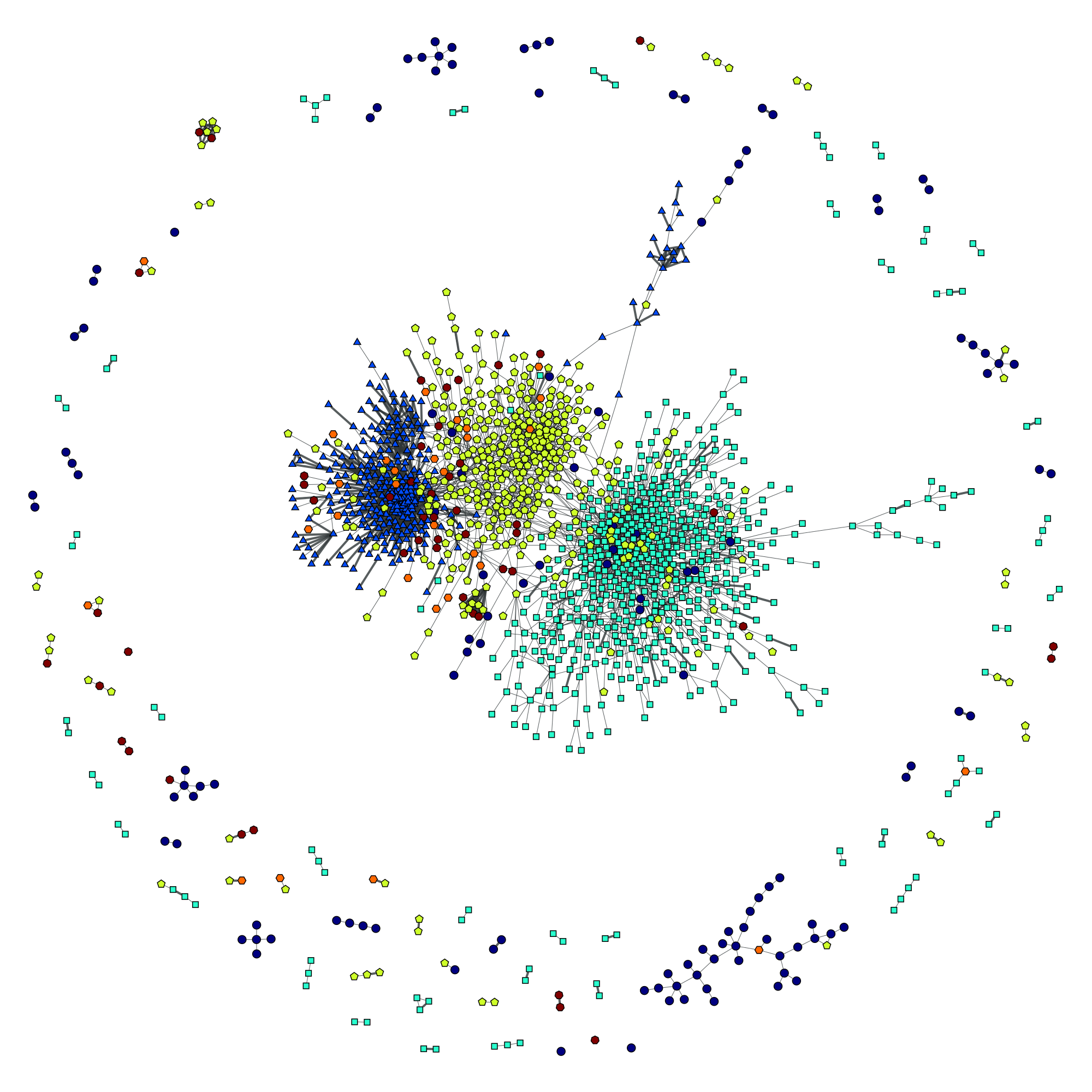}}
\caption{Friendship relations between likers of different campaigns.}
\label{fig:friendshipGraph_withALMS}
\end{figure*}

To evaluate the social ties between likers, we looked at friendship relations between likers (either originating from the same campaign provider or not), ignoring friendship relations with Facebook users who did not like any of our pages. Table~\ref{tab:LikersAndFriendships} (sixth column) reports, for each provider, the overall number of friendship relationships between likers that involved users associated with the provider.

In Figure~\ref{fig:friendship}, we plot the social graph induced by such friendship relations (likers who did not have friendship relations with any other likers were excluded from the graph). %
Based on the resulting social structure, we suggest that: %
\begin{enumerate}
\item
 Dense relations between likers from BoostLikes point to an interconnected network of real users, or fake users who mimic complex ties to pose as real users;
\item
 The pairs (and occasionally triplets) that characterize SocialFormula likers might indicate a different strategy of constructing fake networks, mitigating the risk that identification of a user as fake would consequently bring down the whole connected network of fake users; and
\item
 The friendship relations between AuthenticLikes and MammothSocials likers might indicate that the same operator manages both services.
\end{enumerate}

We also considered indirect links between likers, through mutual friends.
Table~\ref{tab:LikersAndFriendships} reports the overall number of 2-hop relationships between likers from the associated provider.
Figure~\ref{fig:2hop_friendship} plots the relations between likers who either have a direct  relation or a mutual friend, clearly pointing to the presence of relations between likers from the same provider.
These tight connections, along with the number of their friends, suggest that we only see a small part of these networks.
For SocialFormula, AuthenticLikes, and MammothSocials, we also observe many isolated pairs and triplets of likers who are not connected. One possible explanation is that farm users create fake Facebook accounts and keep them separate from their personal accounts and friends.
In contrast, the BoostLikes network is well-connected.

\begin{figure*}[!t]
\centering
         \subfigure[]{\includegraphics[width=0.74\columnwidth]{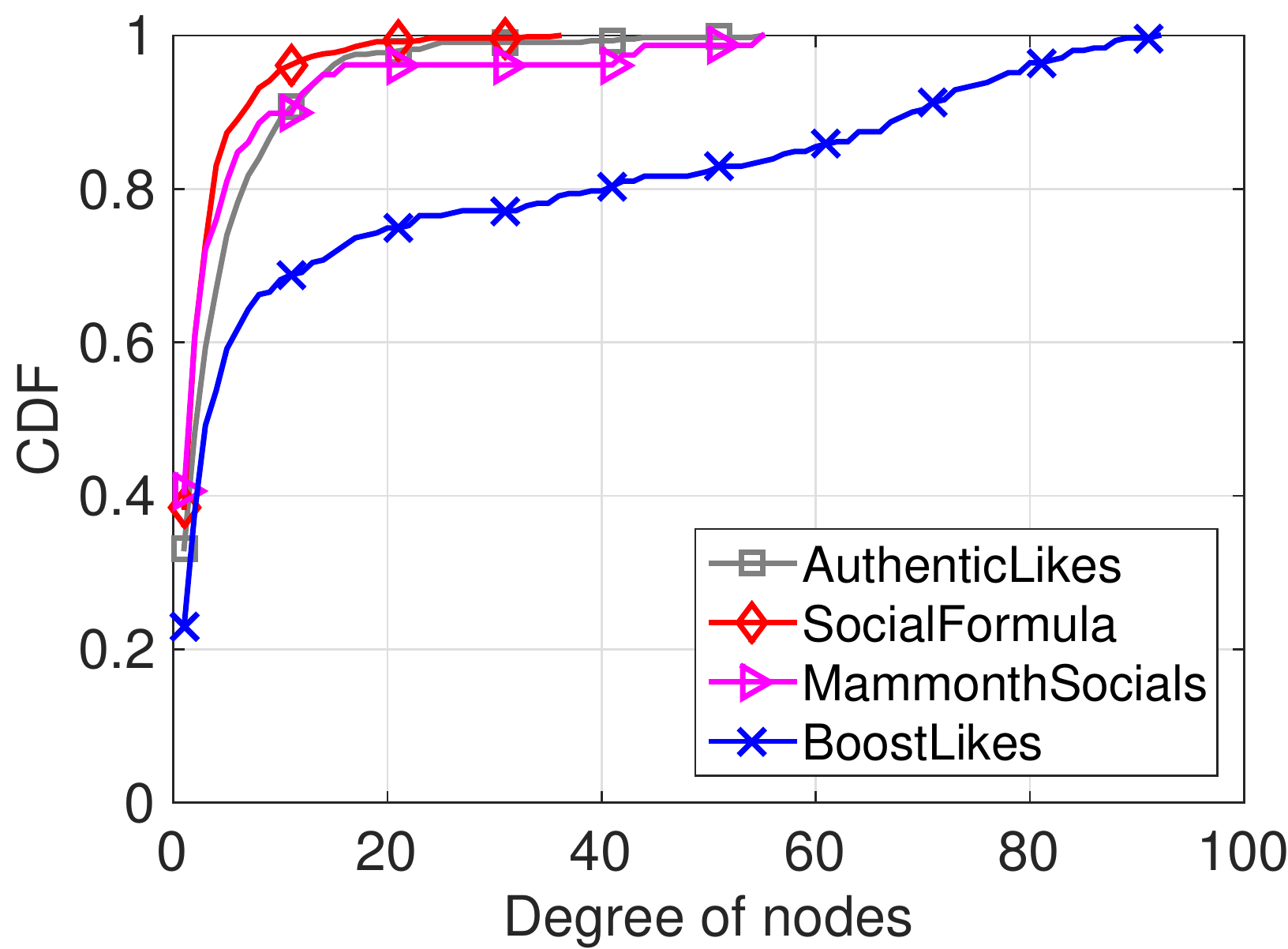}
         \label{Degree distribution}}
         \subfigure[]{\includegraphics[width=0.74\columnwidth]{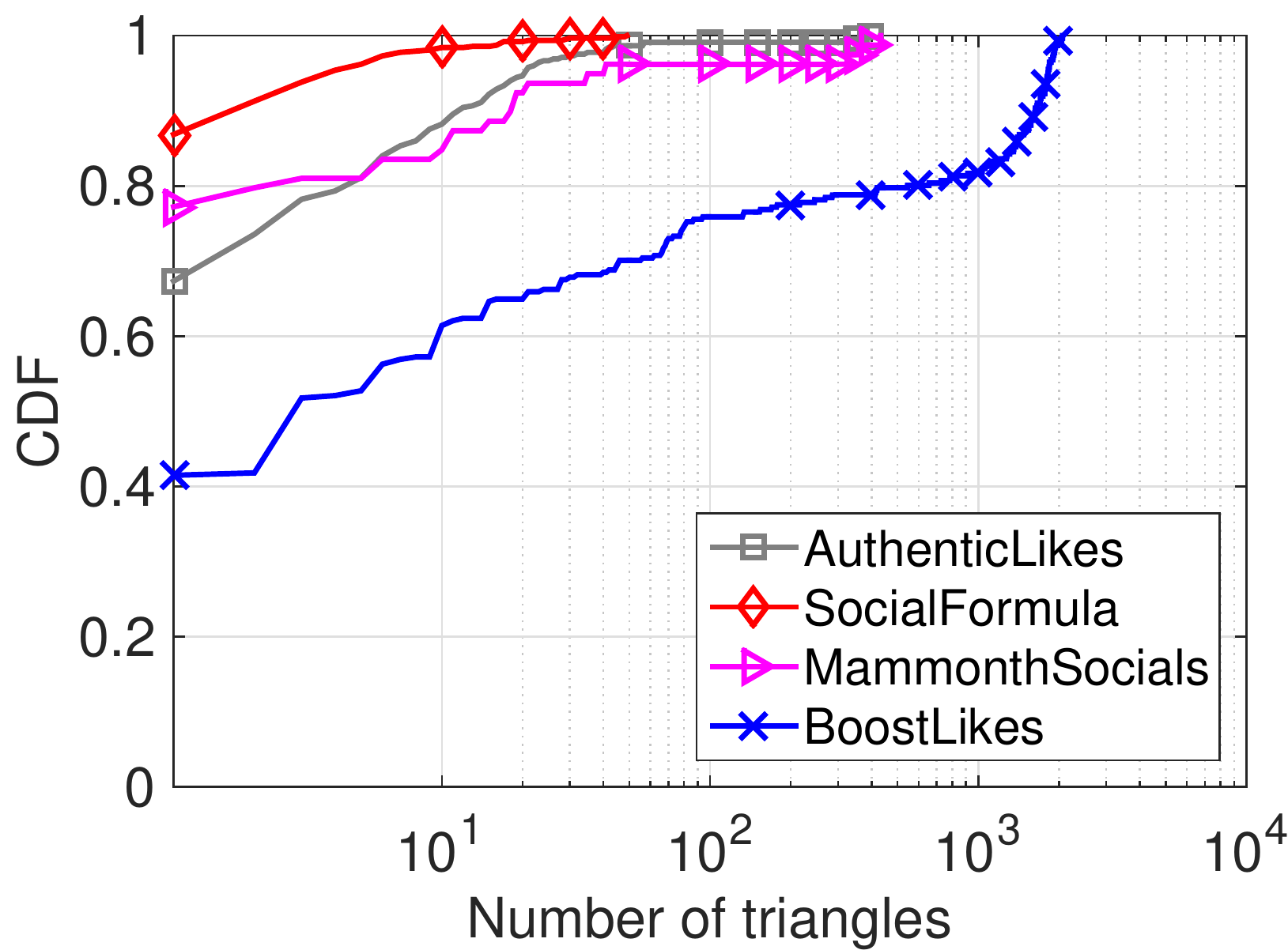}
         \label{Number of triangles}}
	\subfigure[]{\includegraphics[width=0.74\columnwidth]{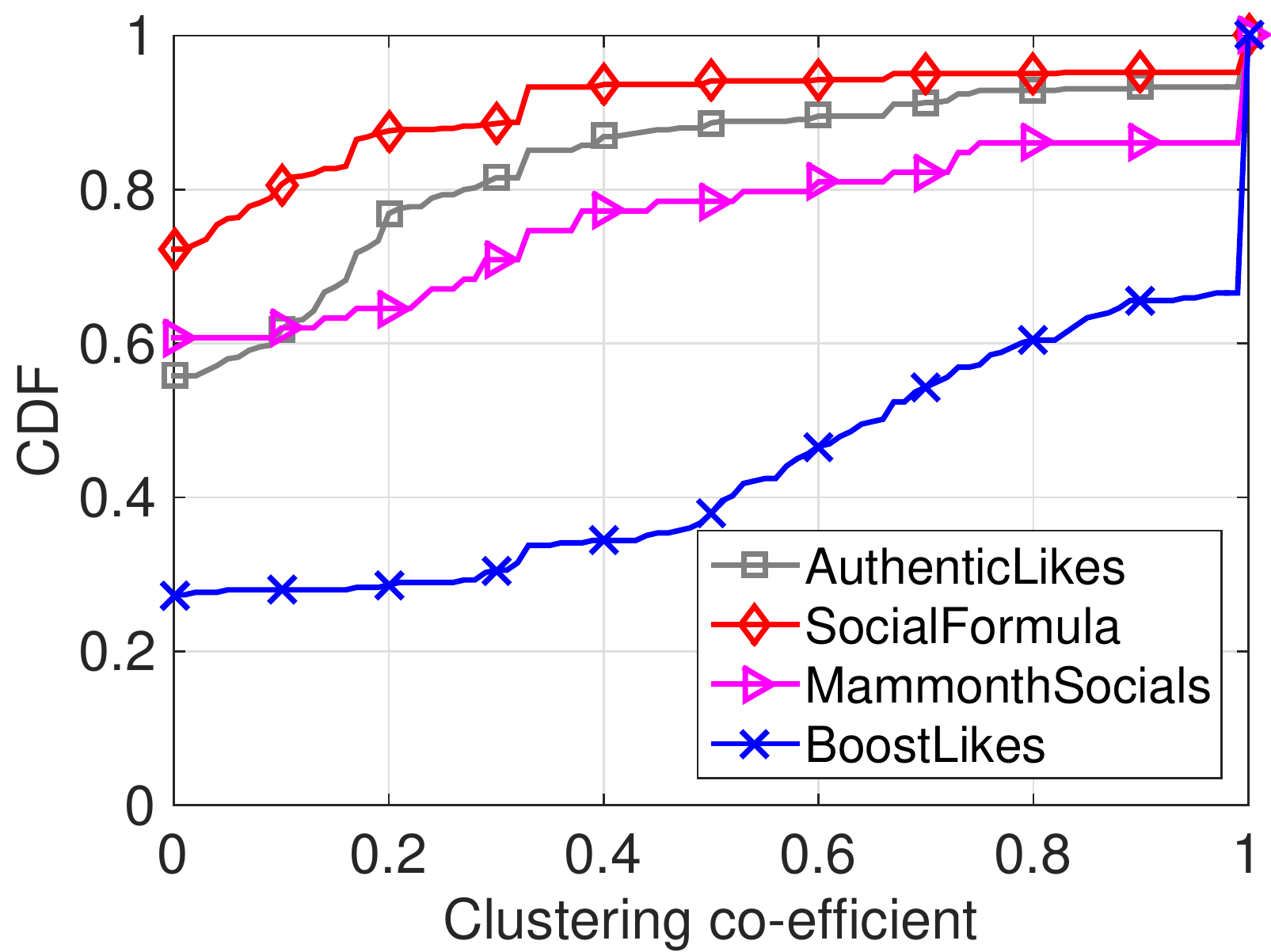}
         \label{Clustering coefficient}}
      \subfigure[]{\includegraphics[width=0.74\columnwidth]{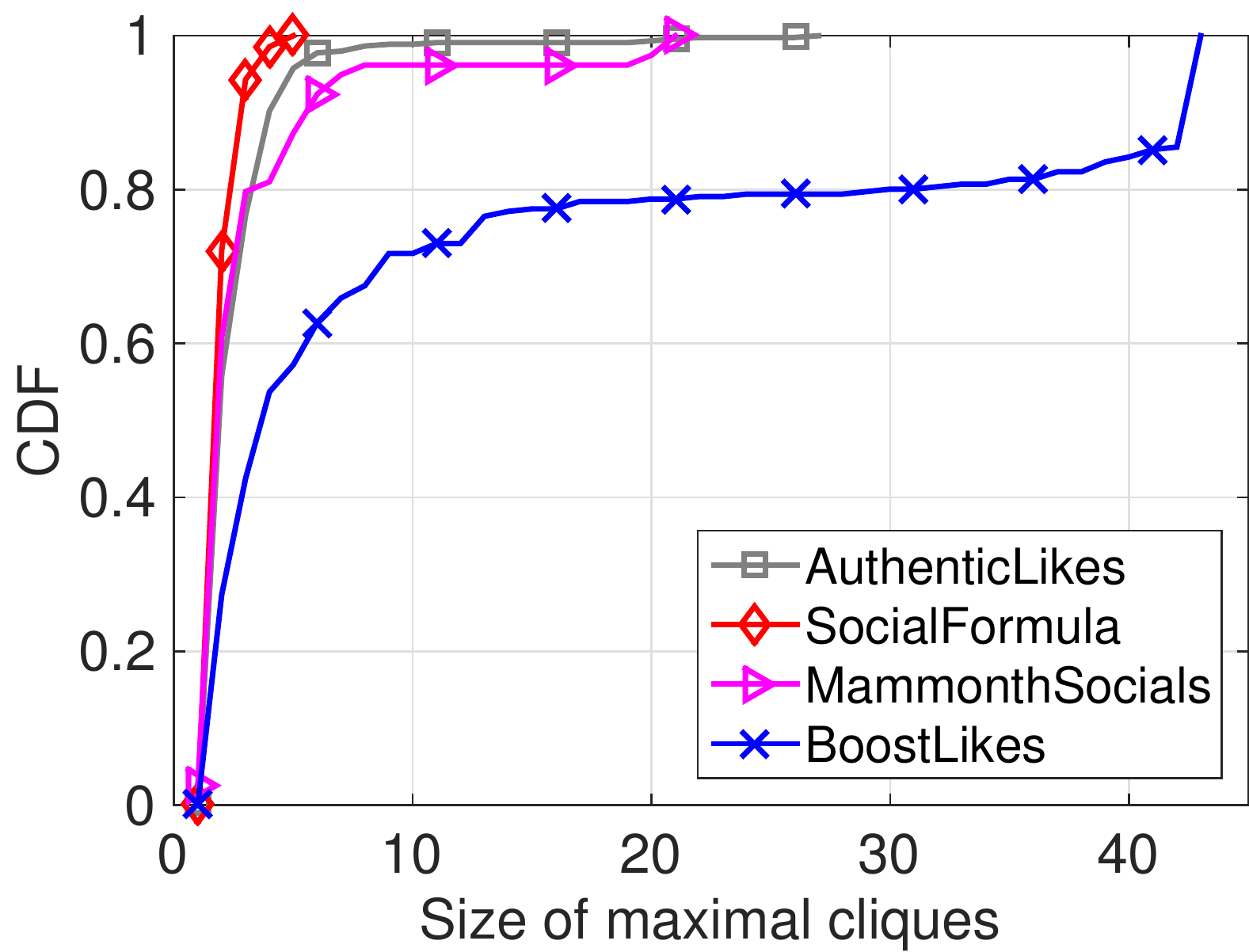}
         \label{Maximal Cliques}}

\caption{Structural properties of the graph of 2-hop relationships among likers of like farm campaigns.}
\label{fig:friendshipGraphProperties}
\end{figure*}

To further compare connectivity of BoostLikes versus SocialFormula, AuthenticLikes, and MammothSocials, we analyze the structural properties of the social graph visualized in Figure \ref{fig:2hop_friendship}.
Figure \ref{fig:friendshipGraphProperties} plots distributions of degree, number of triangles, clustering coefficient, and cliques for these like farms.
The distributions demonstrate that BoostLikes accounts have dense connectivity as compared to accounts belonging to SocialFormula, AuthenticLikes, and MammothSocials.
More specifically, BoostLikes accounts have higher degree, are part of more triangles, have higher clustering coefficient, and have larger maximal cliques than other like farms.
For example, the average degree of BoostLikes accounts is 18 while other like farms have average degrees of less than 5.
Moreover, more than 25\% of BoostLikes accounts make maximal cliques of size greater than 10 while less than 1\% accounts of the other like farms make maximal cliques of size greater than 10.

\subsection{Page Like Analysis}
\label{sec:likeAnalysis}
\begin{figure*}[t!]
\centering
         \subfigure[\small Facebook Campaigns]{\includegraphics[width=0.74\columnwidth]{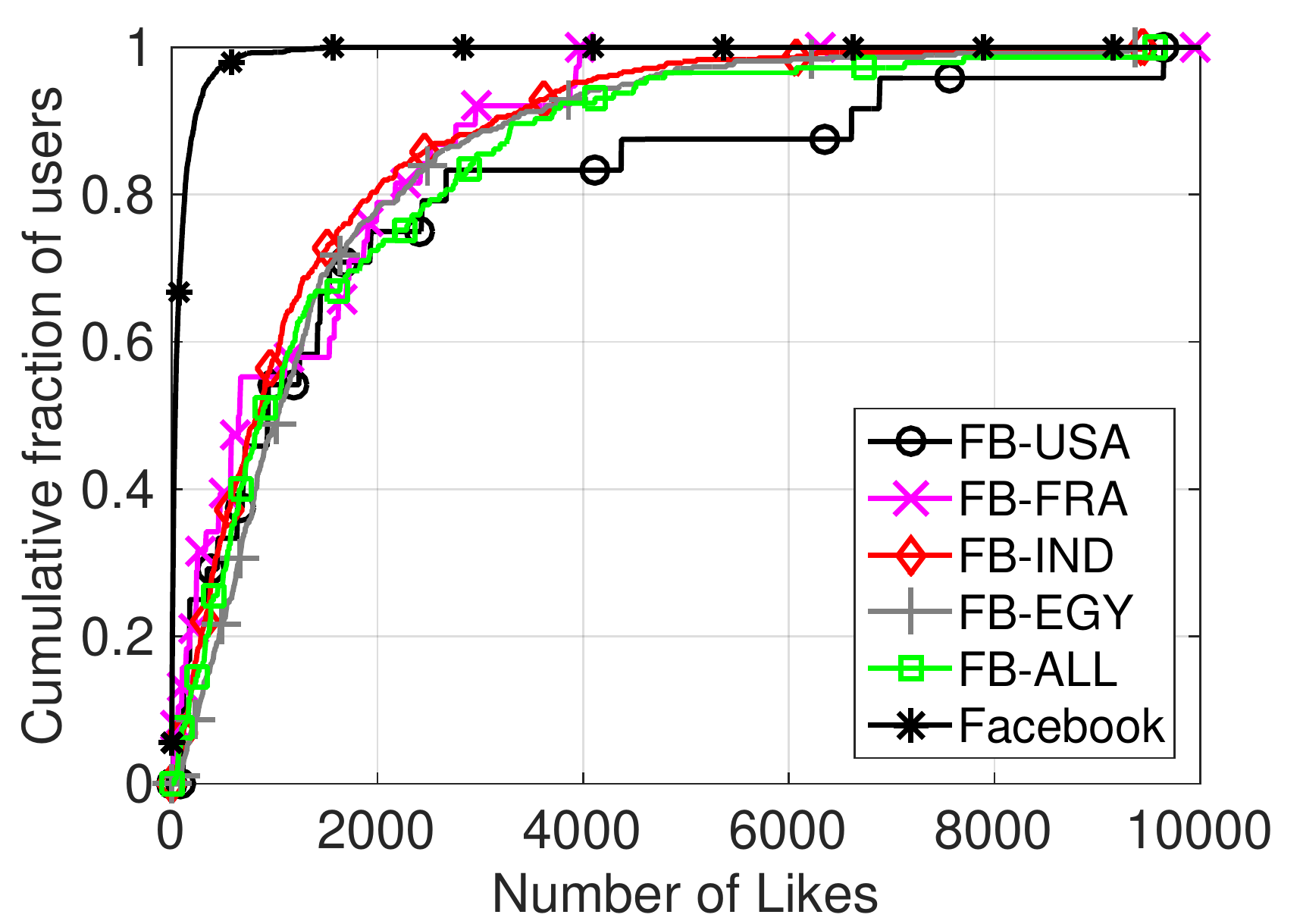}\label{fig:like-distribution-facebook}}
         \subfigure[\small Like Farms]{\includegraphics[width=0.74\columnwidth]{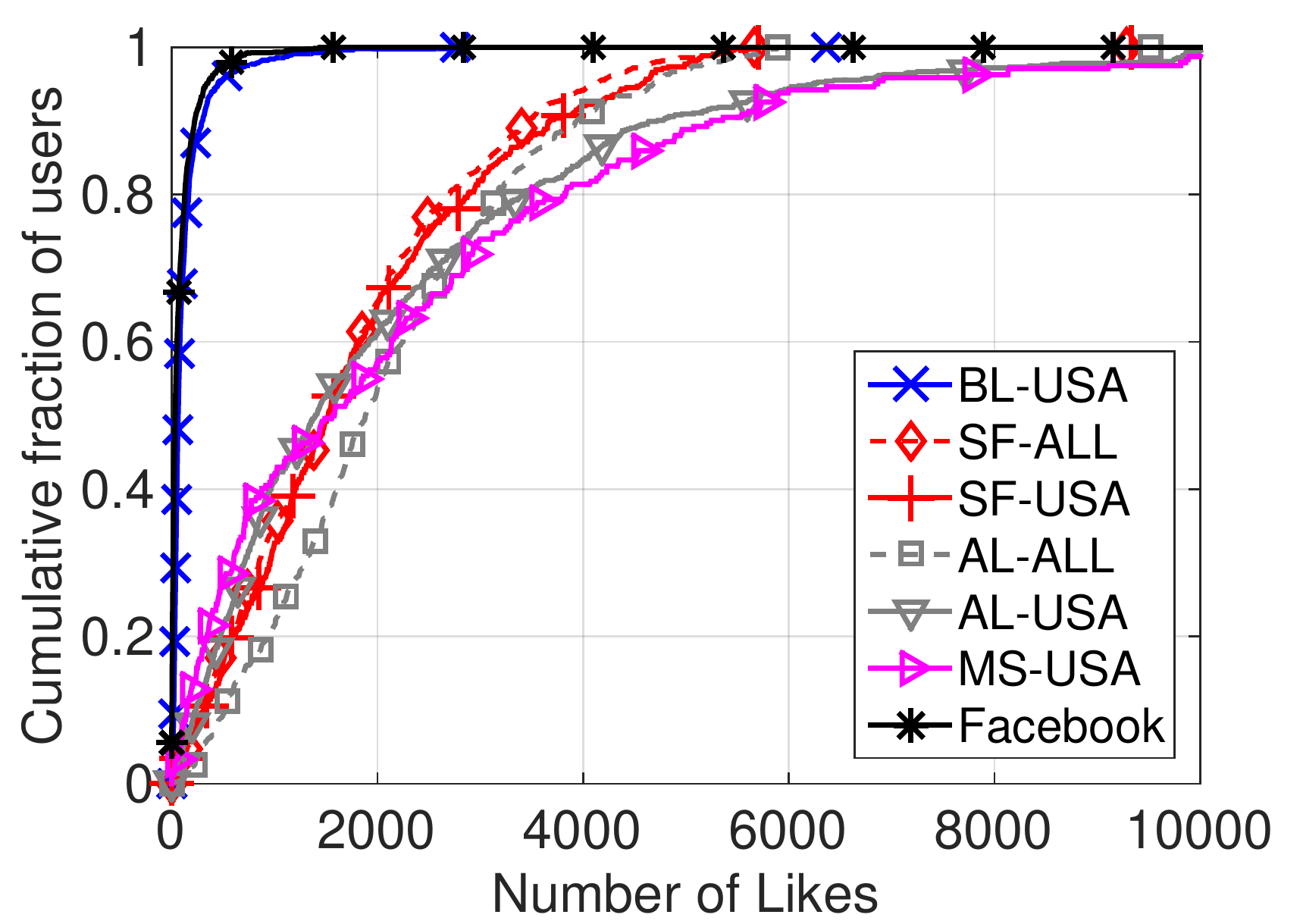}\label{fig:like-distribution-farms}}
\vspace{-0.15cm}
\caption{Distribution of the number of likes by users in Facebook and like farm campaigns.}%
\label{fig:like-distribution}
\end{figure*}

We then looked at the {\em other} pages liked by profiles attracted to our honeypot pages.
In Figure \ref{fig:like-distribution-facebook} and \ref{fig:like-distribution-farms}, respectively,
we plot the distribution of the number of page likes for Facebook ads' and like farm campaigns' users.
To draw a baseline comparison, we also collected page like counts from a random set of 2,000 Facebook users, extracted from an unbiased sample of Facebook user population. The original sample was crawled for another project~\cite{chen2013much}, obtained by randomly sampling Facebook public directory which lists all the IDs of searchable profiles.

We observed a large variance in the number of pages liked, ranging from 1 to 10,000.
The median page like count ranged between 600 and 1000 for users from the Facebook campaigns and between 1200 and 1800 for those from like farm campaigns, with the exception of the BL-USA campaign (median was 63).
In contrast, the median page like count for our baseline Facebook user sample was 34.
The page like counts of our baseline sample mirrored numbers reported in prior work, e.g.,
according to \cite{allfacebook}, the average number of pages liked by Facebook users amounts to roughly 40.
In other words, our honeypot pages attracted users that tend to like significantly more pages than regular Facebook users.
Since our honeypot pages both for Facebook and like farm campaigns explicitly indicated they were not ``real'', we argue that a vast majority of the garnered likes are fake.
We argue that these users like a large number of pages because they are probably reused for multiple ``jobs'' and also like ``normal'' pages to mimic real users.\footnote{Facebook does not impose any limit on the maximum number of page likes per user.}

\begin{figure*}[t!]
\centering
         \subfigure[\small Page Like]{\includegraphics[width=0.75\columnwidth]{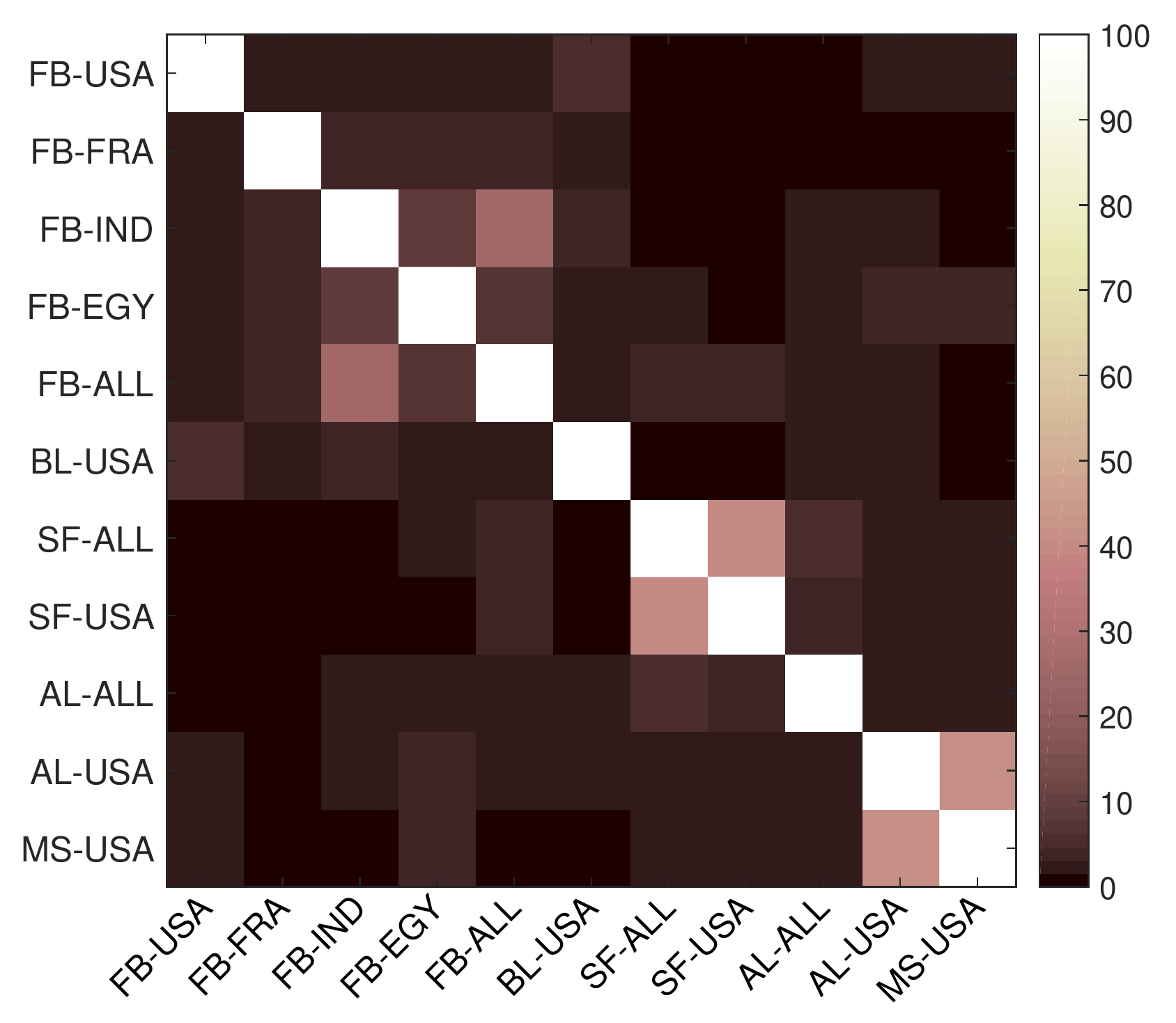}\label{fig:similarity-matrix-a}}
         \subfigure[\small User]{\includegraphics[width=0.75\columnwidth]{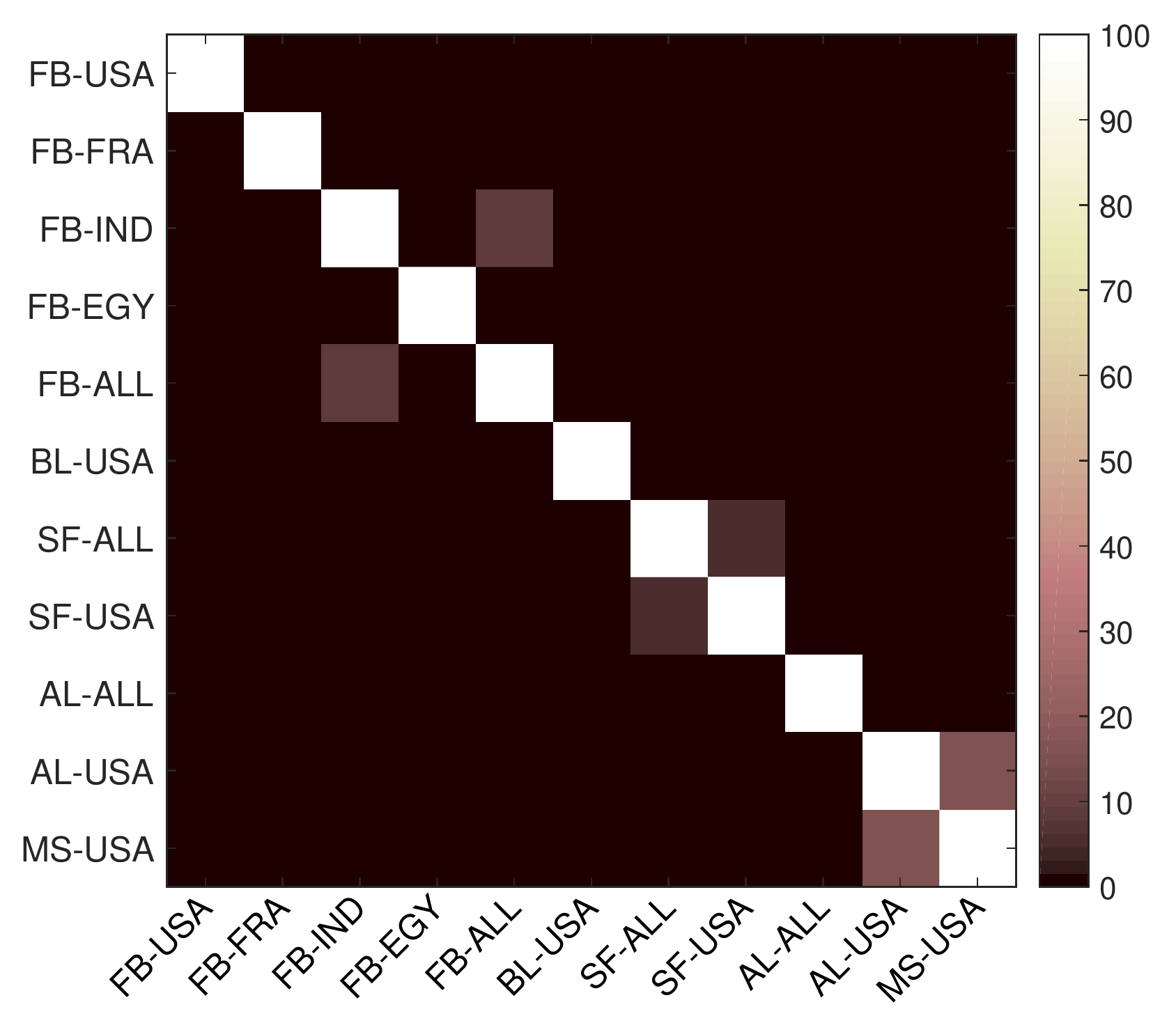}
\label{fig:similarity-matrix-b}}
\vspace{-0.15cm}
\caption{Jaccard index similarity ($\times 100$) matrices of page likes and likers across different campaigns.}
\label{fig:similarity-matrix}
\end{figure*}

To confirm our hypothesis, for each pair of campaigns, we plot their Jaccard similarity.
Specifically, let $S_k$ denote the set of pages liked by a user $k$: the Jaccard similarity between the set of likes by likers of two campaigns $A$ and $B$, which we plot in Figure~\ref{fig:similarity-matrix-a}, is defined as $|A \cap B|/| A \cup B|$,
where $A = \bigcup_{\forall i \in A} S_i $ and $ B = \bigcup_{\forall j \in B} S_j $. %
We also plot, in Figure~\ref{fig:similarity-matrix-b}, the similarity between
$A' = \bigcup_{\forall i \in A} i $ and $ B' = \bigcup_{\forall j \in B} j $, i.e., the similarity between the set of likers of the different campaigns.

Note from Figure \ref{fig:similarity-matrix} that FB-IND, FB-EGY, and FB-ALL have relatively large (Jaccard) similarity with each other.
In addition, the SF-USA and SF-ALL pair and the AL-USA and MS-USA pair also have relatively large Jaccard similarity.
These findings suggest that the same fake profiles are used in multiple campaigns by a like farm (e.g., SF-ALL and SF-USA). Moreover, some fake profiles seem to be shared by different like farms (e.g., AL-USA and MS-USA), suggesting that they are run by the same operator.

\subsection{Discussion}
Overall, we identified two main {\em modi operandi}: (1) some farms, like SocialFormula and AuthenticLikes, seem to be operated by bots and do not really try to hide the nature of their operations, as demonstrated by large bursts of likes and the limited number of friends per profile; (2) other farms, like BoostLikes, follow a much stealthier approach, aiming to mimic regular users' behavior, and rely on their large and well-connected network structure to disseminate the target likes while keeping a small count of likes per user. For the latter, we also observed a high number of friends per profile and a ``reasonable'' number of likes.

A month after the campaigns, we checked whether or not likers' accounts were still active:
as shown in Table~\ref{tbl:measurements-imc}, only one account associated with BoostLikes was terminated, as opposed to 9, 20, and 44 for the other like farms. 11 accounts from the regular Facebook campaigns were also terminated.
Although occurring not so frequently, the accounts' termination might be indicative of the disposable nature of fake accounts on most like farms, where ``bot-like" patterns are actually easy to detect.
It also mirrors the challenge Facebook is confronted by, with like farms such as BoostLikes that exhibit patterns closely resembling real users' behavior, thus making fake like detection quite difficult.

We stress that our findings do not necessarily imply that advertising on Facebook is ineffective, since our campaigns were specifically designed to avert real users. However, we do provide strong evidence that likers attracted on our honeypot pages, even when using legitimate Facebook campaigns, are significantly different from typical Facebook users, which confirms the concerns about the genuineness of these likes.
We also show that most fake likes exhibit some peculiar characteristics -- including demographics, likes, temporal and social graph patterns -- that can and should be exploited by like fraud detection algorithms.

\section{Limitations of Graph Co-Clustering Techniques}
\label{sec:coclustering}

Aiming to counter fraudulent activities, including like farms, Facebook has recently deployed detection tools such as CopyCatch~\cite{beutel2013copycatch} and SynchroTrap~\cite{cao14synchrotrap}. These tools use graph co-clustering algorithms to detect large groups of malicious accounts that like similar pages around the same time frame. However, as shown in Section~\ref{sec:imc}, some stealthy like farms seem to deliberately modify their behavior in order to avoid synchronized patterns, which might reduce the effectiveness of these detection tools. Specifically, while several farms use a large number of accounts (possibly fake or compromised) liking target pages within a short timespan, some spread likes over longer timespans and onto popular pages aiming to circumvent fraud detection algorithms.
In this section, we analyze the efficacy of state-of-the-art co-clustering algorithms on our dataset of like farm users.

\subsection{Re-Crawling}
\label{sec:data-new}
Our experiments use, as ground truth, the Facebook accounts gathered as part of the honeypot-based
measurement of like farms. Recall (from~Section~\ref{sec:imc}) that we garnered 5,918 likes from 5,616 unique users, specifically, 1,437 unique accounts from Facebook ad campaigns and 4,179 unique accounts from the like farm campaigns (note that some users liked more than one honeypot pages).
In Summer 2015, we checked how many accounts had been closed or terminated and found that 624 out of 5,616 accounts (11\%) were no longer active. We then began to crawl the pages liked by each of the 4,179 like farm users (again, using Selenium web driver). We collected basic information associated with each page, such as the total number of likes, category, and location, using the page identifier. Unlike our previous crawl, we now also collected the timelines of the like farm accounts, specifically, timeline posts (up to a maximum of 500 recent posts), the comments on each post, as well as the associated number of likes and comments on each post.

Besides some accounts having become inactive (376), we also could not crawl the timeline of 24 users who had restricted the visibility of their timeline. Moreover, in Fall 2015, Facebook blocked all the accounts we were using for crawling, and so we stopped our data collection before we could completely finish our data collection, hence, we missed further 109 users. In summary, our new dataset consists of 3,670 users (out of the initial 4,179), with more than 234K posts (messages, shared content, check-ins, etc) for these accounts. %
In our experiments, we will also rely on a baseline of 1,408 random accounts from Chen et al.~\cite{chen2013much} %
which we use to form a baseline of ``normal'' accounts. For each of these accounts, we again collected posts from their timeline, their page likes, and information from these pages. 53\% of the accounts had at least 10 visible posts on the timeline, and in total we collected about 35K posts.

\begin{table}[t]
\centering
\tabcolsep=0.11cm
\resizebox{0.45\textwidth}{!}{%
\begin{tabular}{lrrrr}
\toprule
{\bf Campaign} & {\bf \#Users} & {\bf \#Pages} &  {\bf \#Pages Liked} & {\bf \#Posts}  \\
&  & {\bf Liked} & {\bf (Unique)} & \\
\midrule
BL-USA & 583 & 79,025 & 37,283 & 44,566\\ %
SF-ALL & 870 & 879,369 & 108,020 &46,394 \\ %
SF-USA  & 653 & 340,964 & 75,404 & 38,999 \\ %
AL-ALL  & 707 & 162,686 & 46,230 &61,575\\  %
AL-USA   & 827 & 441,187 & 141,214 & 30,715\\ %
MS-USA  & 259 & 412,258 & 141,262 & 12,280\\  \hline %
{\em Tot. Farms} & {\em 3,899} & {\em 2,315,489} & {\em 549,413}  & {\em 234,529}\\
\midrule
{Baseline}  & 1,408 & 79,247 & 57,384 & 34,903\\
\bottomrule
\end{tabular}
}
\vspace{-0.15cm}
\caption{Overview of the datasets used in our study.}
\label{tbl:measurements-new}
\end{table}

Table~\ref{tbl:measurements-new} summarizes the data used in the experiments presented in the rest of the paper. Note that users who like more than one honeypot pages are included in all rows, hence the disparity between the number of unique users (3,670) and the total reported in the table (3,899). Overall, we gathered information from 600K unique pages, liked by 3,670 like farm accounts and 1,408 baseline accounts, and around 270K  posts.

Again, note that we collected openly available data such as (public) profile and timeline information, as well as page likes. Also, all data was encrypted at rest and has not been re-distributed. No personal information was extracted as we only analyzed aggregated statistics. We also consulted Data61's legal team, which classified our research as exempt and likewise, received approval from the ethics committee of UCL.

\subsection{Experimental Evaluation of Co-Clustering}

We use the labeled dataset of 3,670 users from six different like farms and the 1,408 baseline users, and employ a graph co-clustering algorithm to divide the user-page bipartite graph into distinct clusters~\cite{Kluger03biclustering}. Similar to CopyCatch~\cite{beutel2013copycatch} and SynchroTrap~\cite{cao14synchrotrap}, the clusters identified in the user-page bipartite graph represent near-bipartite cores, and the set of users in a near-bipartite core like the same set of pages. Since we are interested in distinguishing between two classes of users (like farm users and normal users), we set the target number of clusters at 2. Given that our crawlers were restricted to crawl the behavior of all like farms and baseline users on daily basis, we do not have fine-grained features to further analyze CopyCatch and SynchroTrap. Aiming to reveal the liking behavior of like farms users, we evaluate the employed graph co-clustering schemes of CopyCatch and SynchroTrap on our collected datasets. %

\begin{table}[t]
  \begin{center}
\tabcolsep=0.11cm
\resizebox{0.475\textwidth}{!}{%
    \begin{tabular}{lrrrrrrr}
      \toprule
      {\bf Campaign} & {\bf TP} & {\bf FP} & {\bf TN} & {\bf FN} & {\bf Precision} & {\bf Recall} & {\bf F1-Score} \\
      \midrule
AL-USA & 681 & 9 & 569 & 4 & 98\% & 99\% & 99\% \\
AL-ALL & 448 & 53 & 527 & 1 & 89\% & 99\% & 94\% \\
{\bf BL-USA} & 523 & 588 & 18 & 0 & \textbf{47\%} & 100\% & {\bf 64\%} \\
SF-USA & 428 & 67 & 512 & 1 & 86\% & 100\% & 94\% \\
SF-ALL & 431 & 48 & 530 & 2 & 90\% & 99\% & 95\% \\
MS-USA & 201 & 22 & 549 & 2 & 90\% & 99\% & 93\% \\
      \bottomrule
    \end{tabular}
}
\vspace{-0.15cm}
      \caption{Effectiveness of the graph co-clustering algorithm.}
          \label{tab:clustering_accuracy}
  \end{center}
\end{table}

\begin{figure*}[p]
\centering
\subfigure[AL-USA]
{\includegraphics[width=0.85\columnwidth]{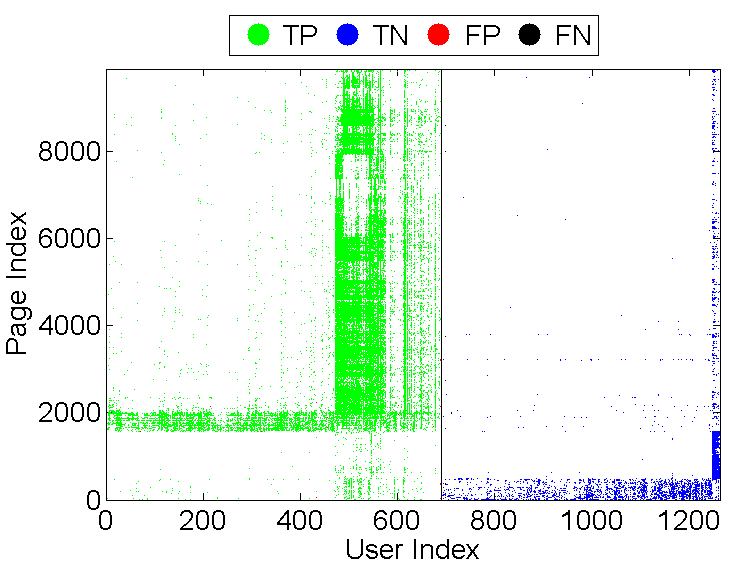}}
~
\subfigure[AL-ALL]
{\includegraphics[width=0.85\columnwidth]{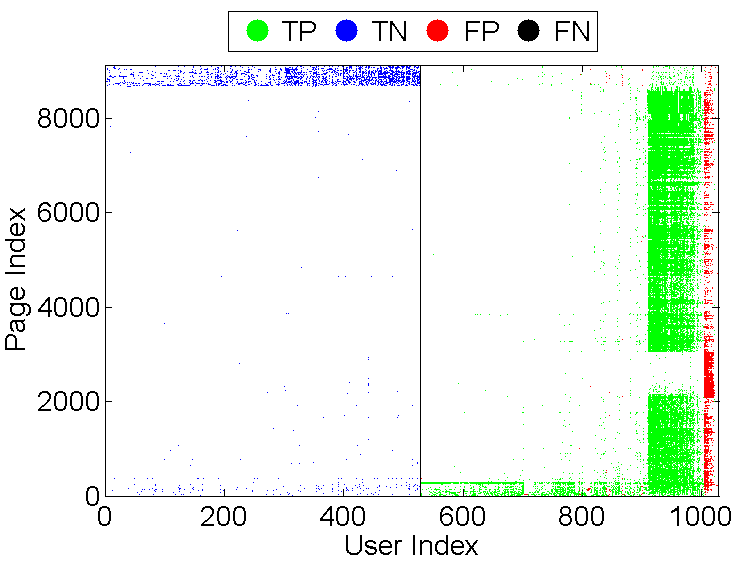}}\\
\subfigure[BL-USA]
{\includegraphics[width=0.85\columnwidth]{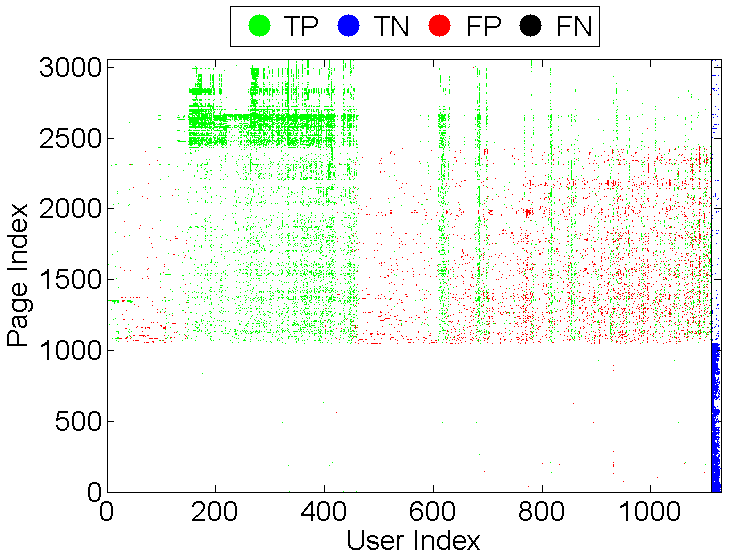}}
~
\subfigure[SF-USA]
{\includegraphics[width=0.85\columnwidth]{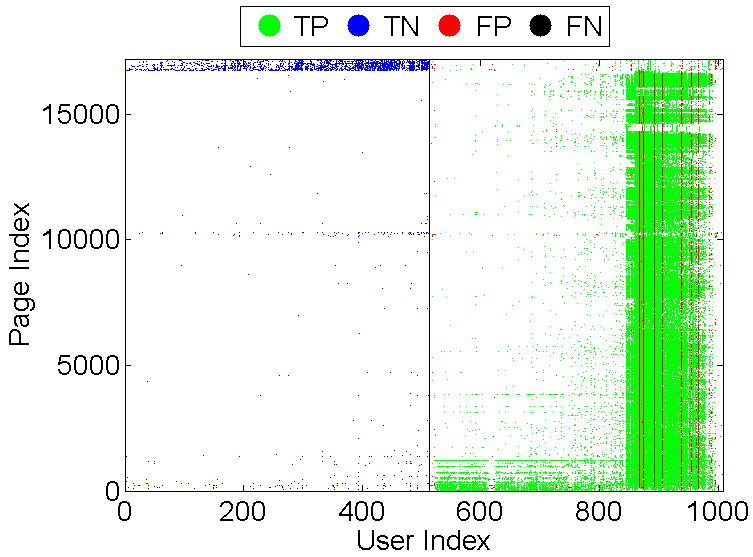}}\\
\subfigure[SF-ALL]
{\includegraphics[width=0.85\columnwidth]{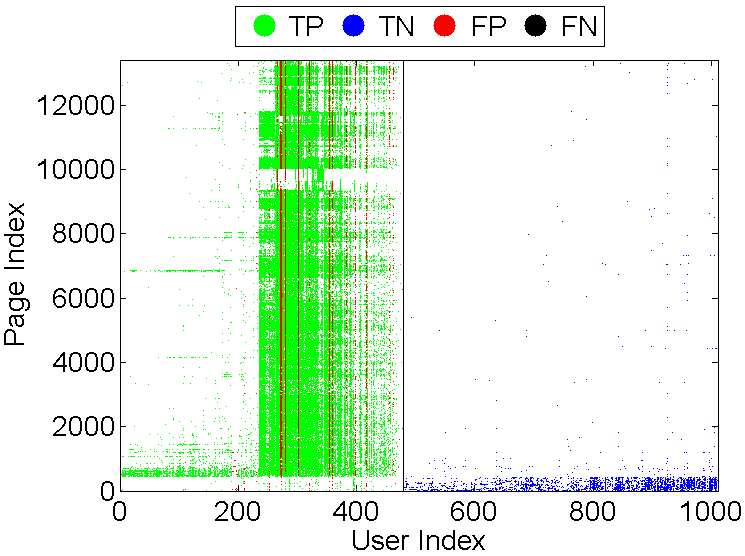}}
~
\subfigure[MS-USA]
{\includegraphics[width=0.85\columnwidth]{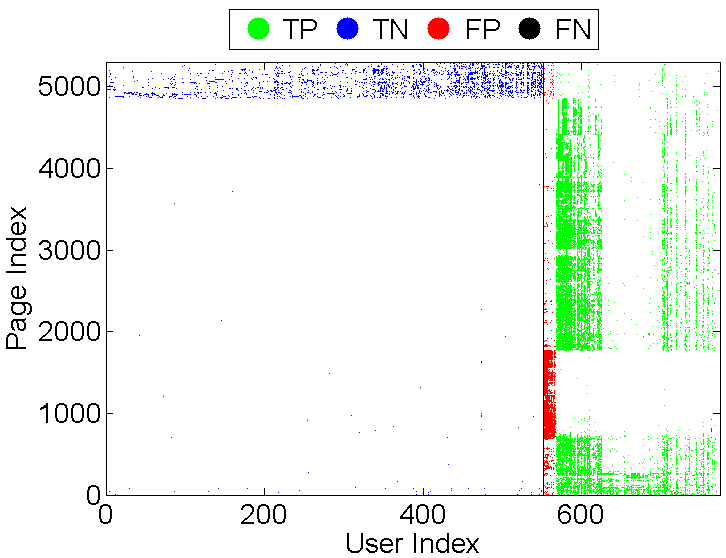}}
\vspace{0.2cm}
\caption{Visualization of graph co-clustering results. The vertical black line indicates the separation between two clusters. We note that the clustering algorithm fails to achieve good separation leading to a large number of false positives (red dots).}
\label{fig: clustering results}
\end{figure*}

\descr{Results.} In Table~\ref{tab:clustering_accuracy}, we report the receiver operating characteristic (ROC) statistics of the graph co-clustering algorithm -- specifically, true positives (TP), false positives (FP), true negatives (TN), false negatives (FN), Precision: $(TP)/(TP+FP)$, Recall: $(TP)/(TP+FN)$, and F1-Score, i.e., the harmonic average of precision and recall. \figurename~\ref{fig: clustering results} visualizes the clustering results as user-page scatter plots. The x-axis represents the user index and the y-axis the page index.\footnote{To ease presentation, we exclude users and pages with less than 10 likes.} The vertical black line marks the separation between two clusters. The points in the scatter plot are colored to indicate true positives (green), true negatives (blue), false positives (red), and false negatives (black).

\descr{Analysis.} We observe two distinct behaviors in the scatter plots: (1) {\em ``liking everything''} (vertical streaks), and (2) {\em ``everyone liking a particular page''} (horizontal streaks). Both like farm and normal users exhibit vertical and horizontal streaks in the scatter plots.

While the graph co-clustering algorithm neatly separates users for AL-USA, it incurs false positives for other like farms. In particular, the co-clustering algorithm fails to achieve a good separation for BL-USA, where it incurs a large number of false positives, resulting in 47\% precision. Further analysis reveals that the horizontal false positive streaks in BL-USA include popular pages, such as ``Fast \& Furious" and ``SpongeBob SquarePants,'' each with millions of likes. We deduce that stealthy like farms, such as BL-USA, use the tactic of liking popular pages aiming to mimic normal users, which reduces the accuracy of the graph co-clustering algorithm.

Our results highlight the limitations of prior graph co-clustering algorithms in detecting fake likes by like farm accounts. We argue that fake liking activity is challenging to detect when only relying on monitoring the liking activity due to the increased sophistication of stealthier like farms. Therefore, as we discuss next, we plan to leverage the characteristics of timeline features to improve accuracy.

\begin{figure*}[t]
	\centering
\subfigure[]{\includegraphics[width=0.74\columnwidth]{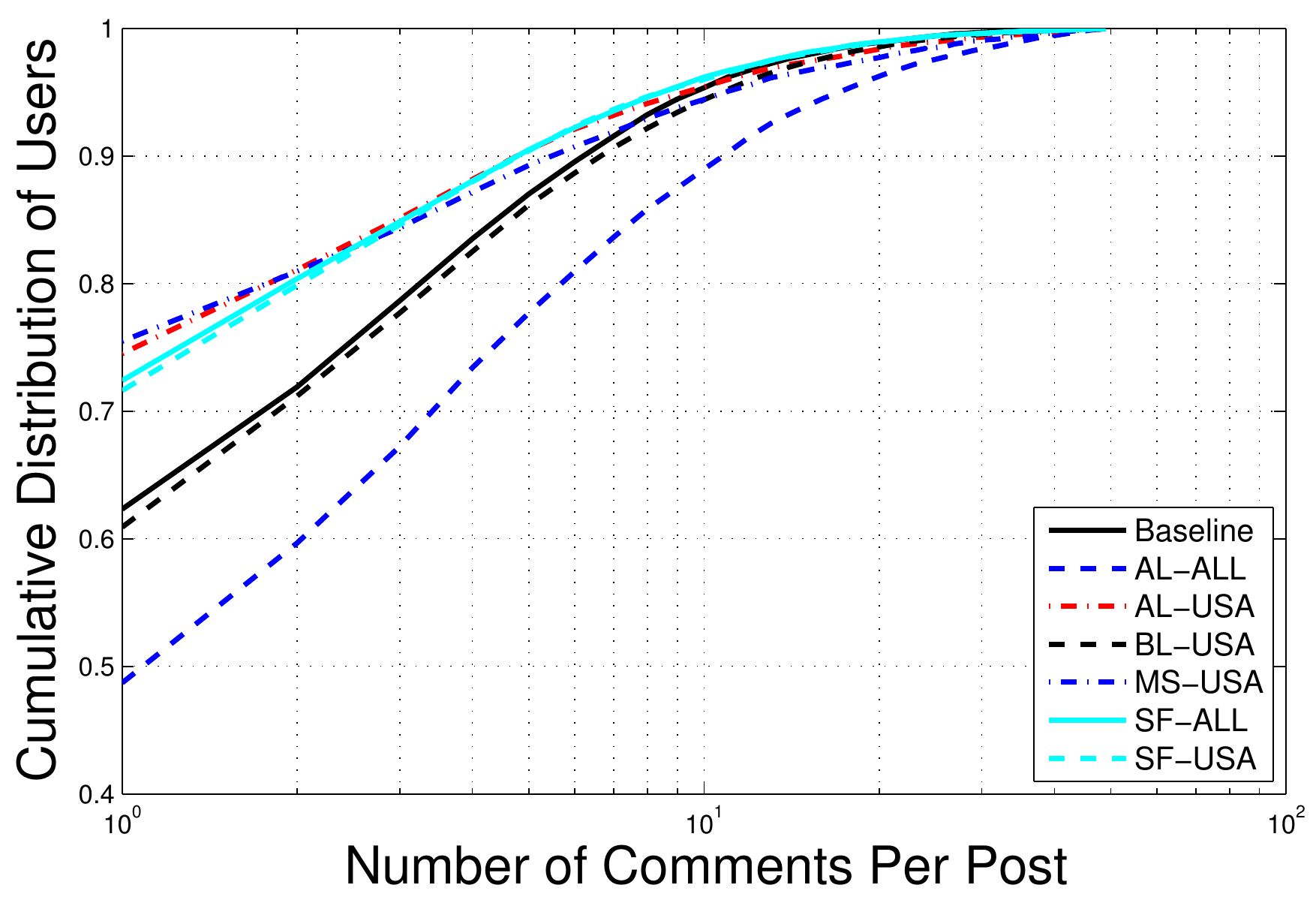}\label{fig:comperpost}}
~\hspace{-0.2cm}
\subfigure[]{\includegraphics[width=0.74\columnwidth]{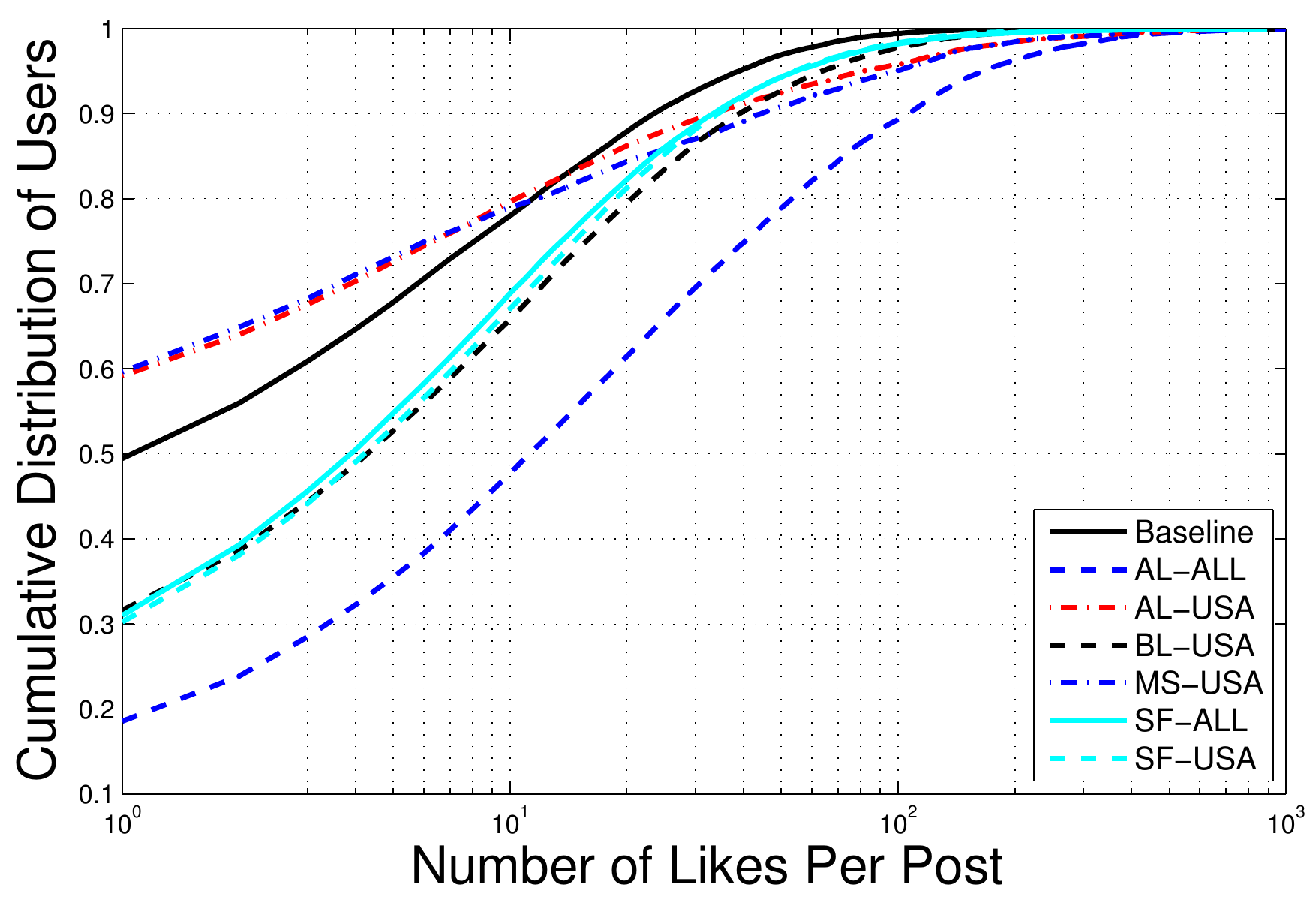}\label{fig:likes-post}}
\\
\subfigure[]{\includegraphics[width=0.74\columnwidth]{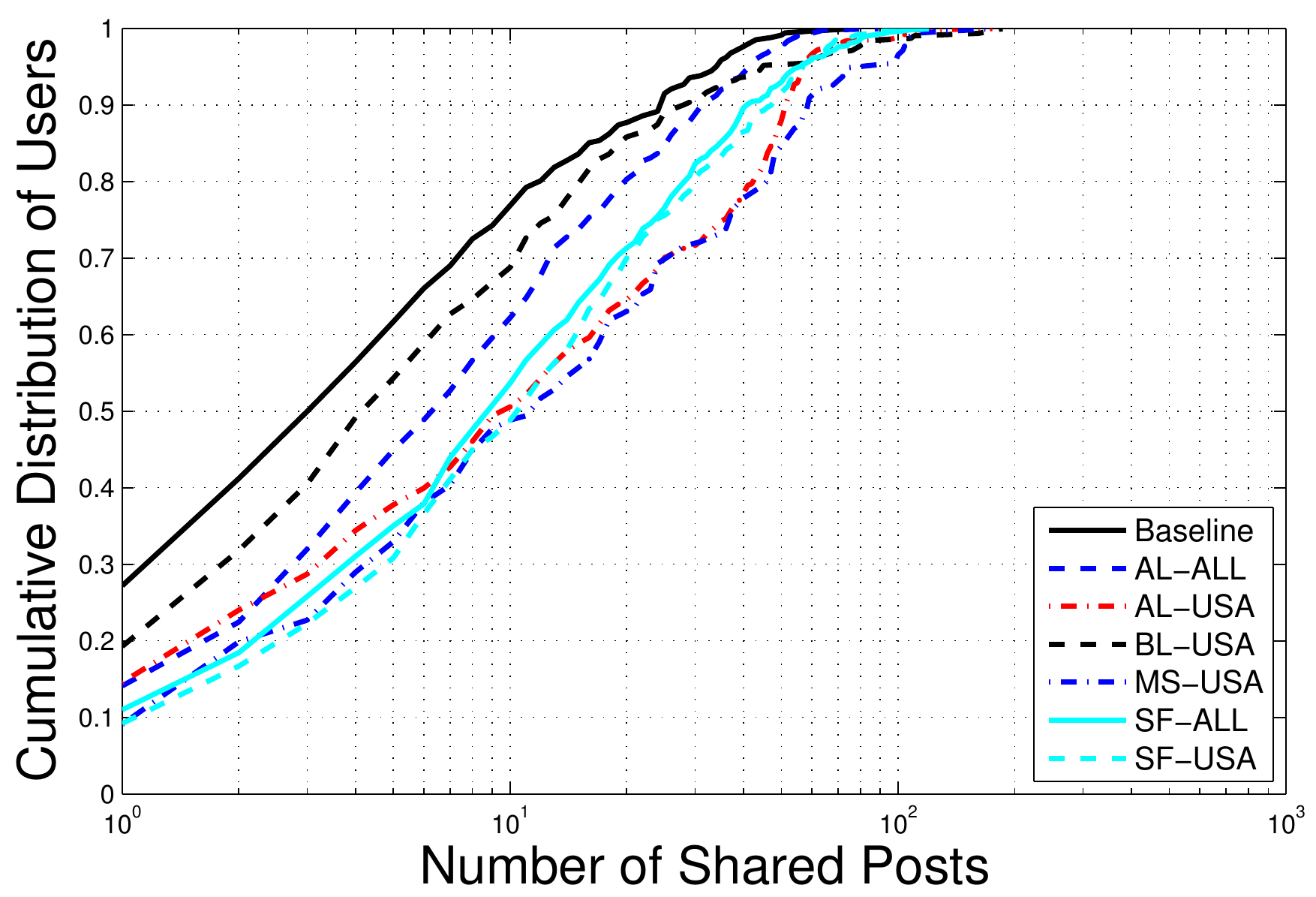}\label{fig:share}}
~\hspace{-0.2cm}
\subfigure[]{\includegraphics[width=0.74\columnwidth]{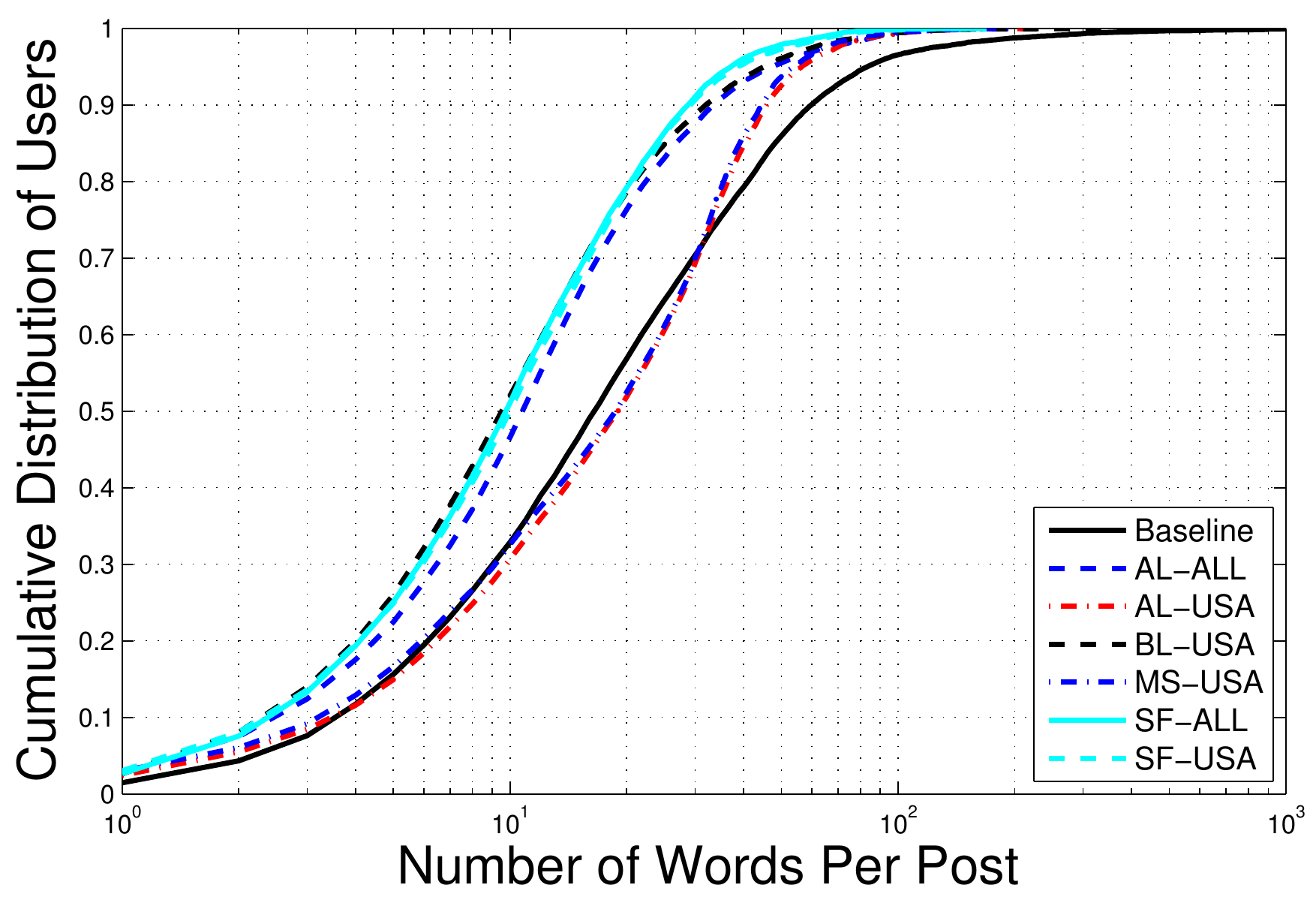}\label{fig:text}}
\vspace{-0.15cm}
\caption{Distribution of non-lexical features for like farm and baseline accounts.}
\end{figure*}

\section{Characterizing Timeline Features}
\label{sec:characterizing}
Motivated by the poor accuracy of graph co-clustering based detection tools on stealthy farms, we set to evaluate the feasibility of timeline-based detection of like farm accounts. To this end, we characterize timeline activities for users in our datasets (cf.~Section~\ref{sec:data-new}) with respect to two categories of features, {\em non-lexical} and {\em lexical}, aiming to identify the most distinguishing features to be used by machine learning algorithms (in Section~\ref{sec:detection}) for accurately classifying like farms vs regular accounts.

\subsection{Analysis of Non-Lexical Features}
\label{sec:charNL}

\descr{Comments and Likes.} In \figurename~\ref{fig:comperpost}, we plot the distributions of the number of comments a post attracts, revealing that users of AL-ALL like farm generate many more comments than the baseline users. We note that BL-USA is almost identical to the baseline users. Next, \figurename~\ref{fig:likes-post} shows the number of likes associated with users' posts, highlighting that posts of like farm users attract much more likes than those of baseline users. Therefore, posts produced by the former gather more likes (and also have lower lexical richness as shown later on in Table~\ref{tab:lexicalfeatures}), which might actually indicate their attempt to mask suspicious activities.

\descr{Shared Content.} We next study the distributions of posts that are classified as ``shared activity," i.e.,  originally made by another user, or articles, images, or videos linked from an external URL (e.g., a blog or YouTube). \figurename~\ref{fig:share} shows that baseline users generate more original posts, and share fewer posts or links, compared to farm users.

\descr{Words per Post.}
\figurename~\ref{fig:text} plots the distributions of number of words that make up a text-based post, highlighting that posts of like farm users tend to have fewer words. Roughly half of the users in four of the like farms (AL-ALL, BL-USA, SF-ALL, and SF-USA) use 10 or less words in their posts, versus 17 words by baseline users.

\begin{figure*}[t]
	\centering
	\subfigure[AL]{
		\includegraphics[width=.35\textwidth]{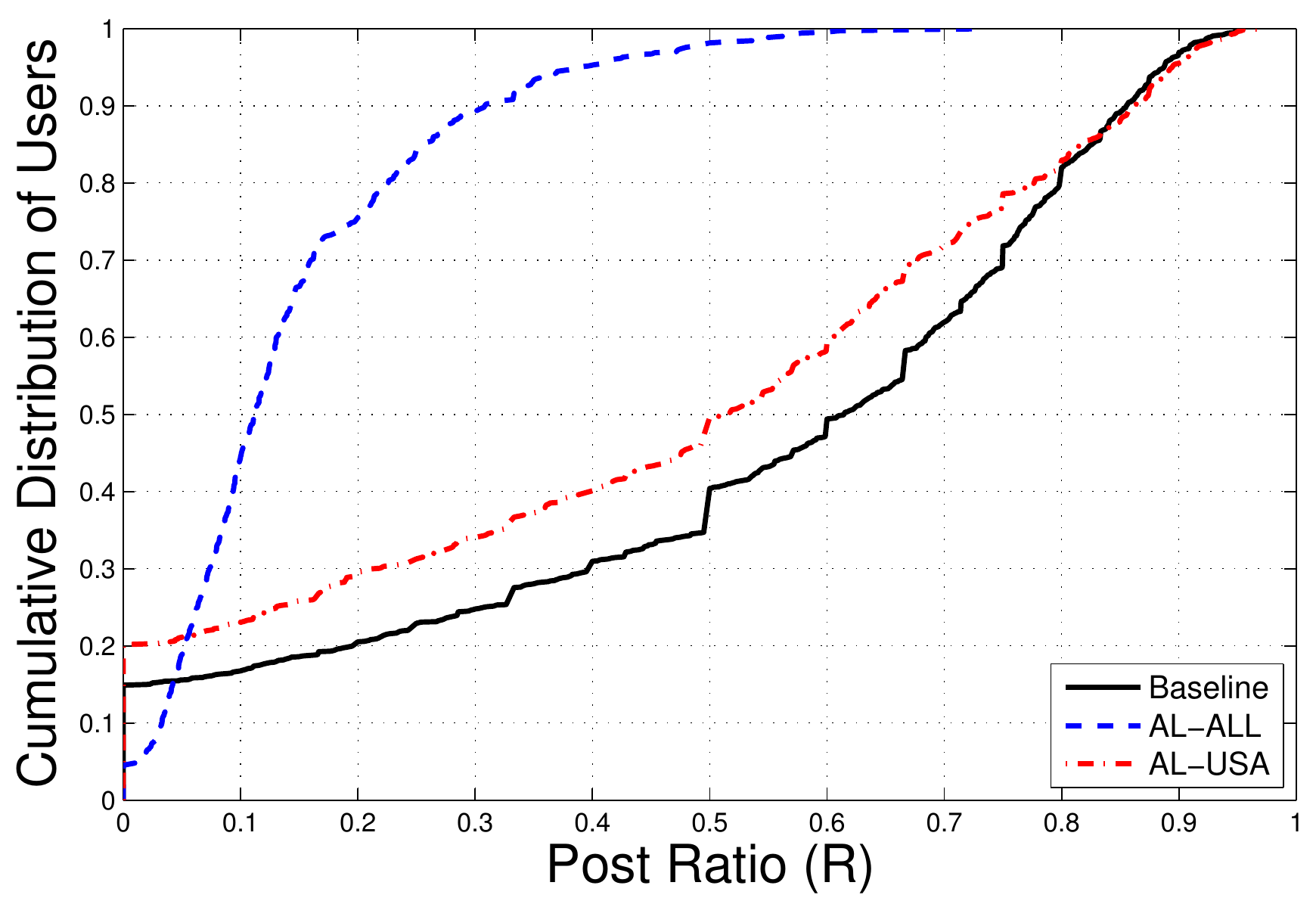}
		\label{fig:al_campaign}
	}\hspace{-0.5cm}
	~
	\subfigure[BL]{
		\includegraphics[width=.35\textwidth]{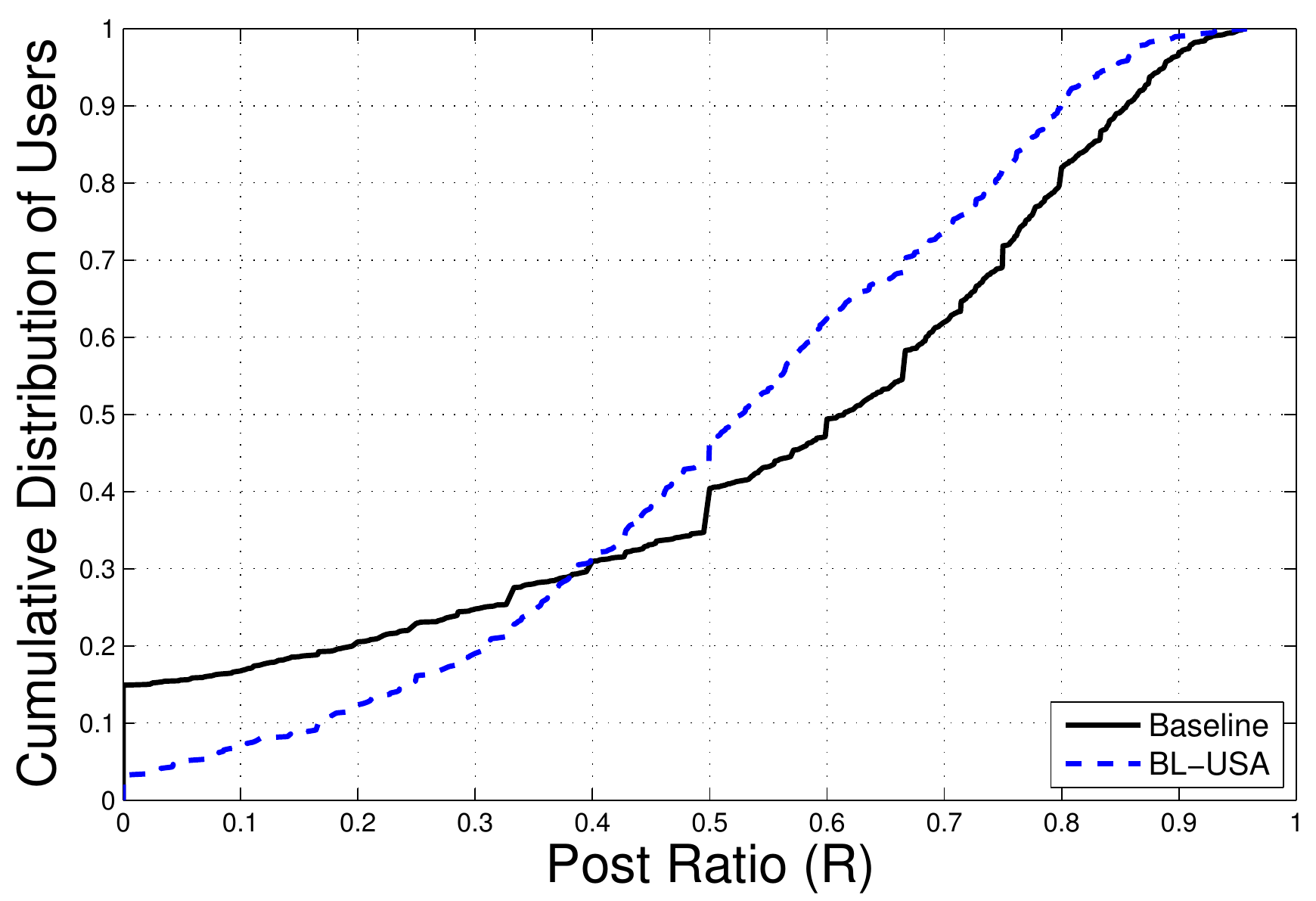}
		\label{fig:bl_campaign}
	}
	\\
	\subfigure[SF]{
		\includegraphics[width=.35\textwidth]{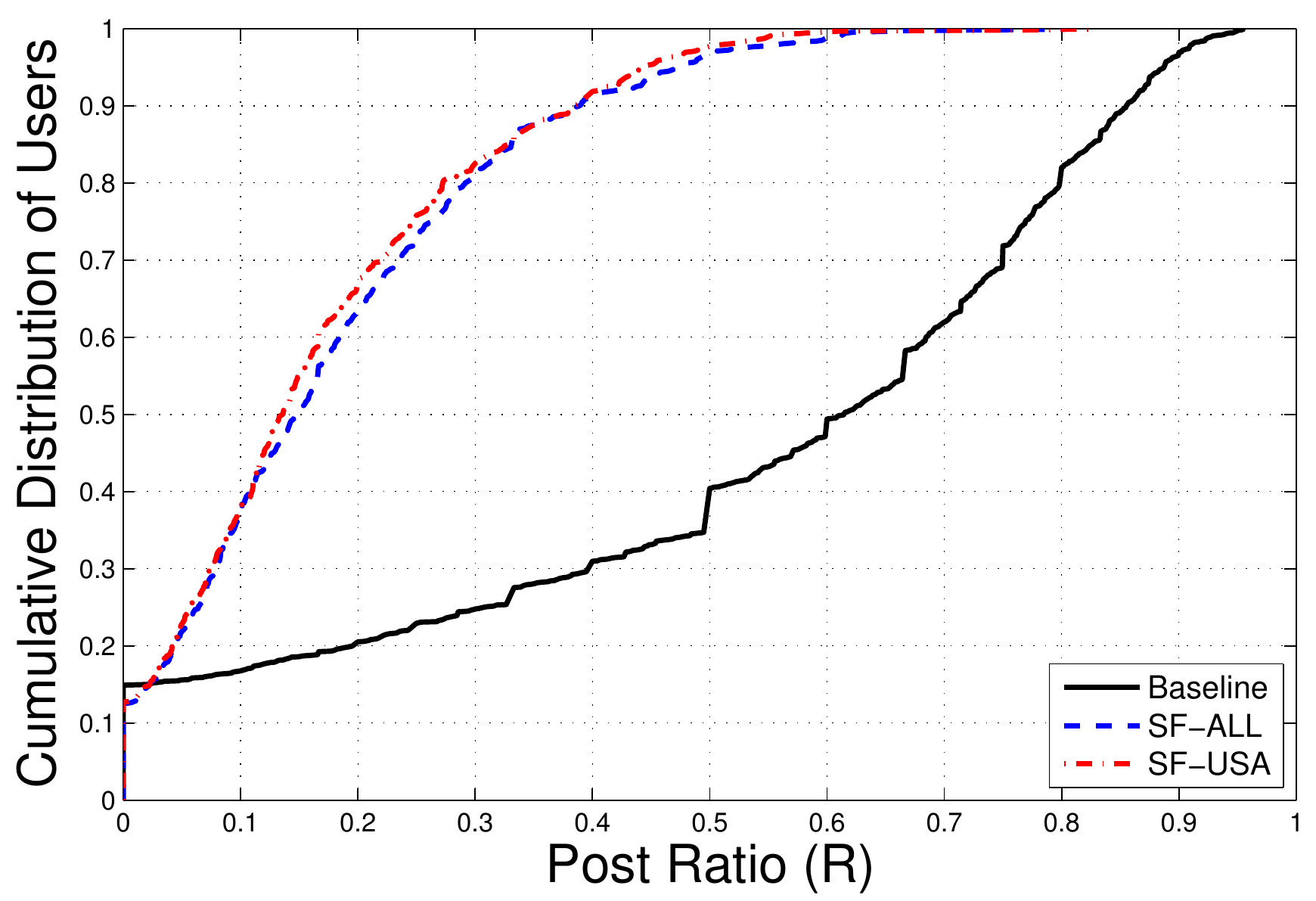}
		\label{fig:ms_us_n_sf}
	}\hspace{-0.5cm}
	~
	\subfigure[MS]{
		\includegraphics[width=.35\textwidth]{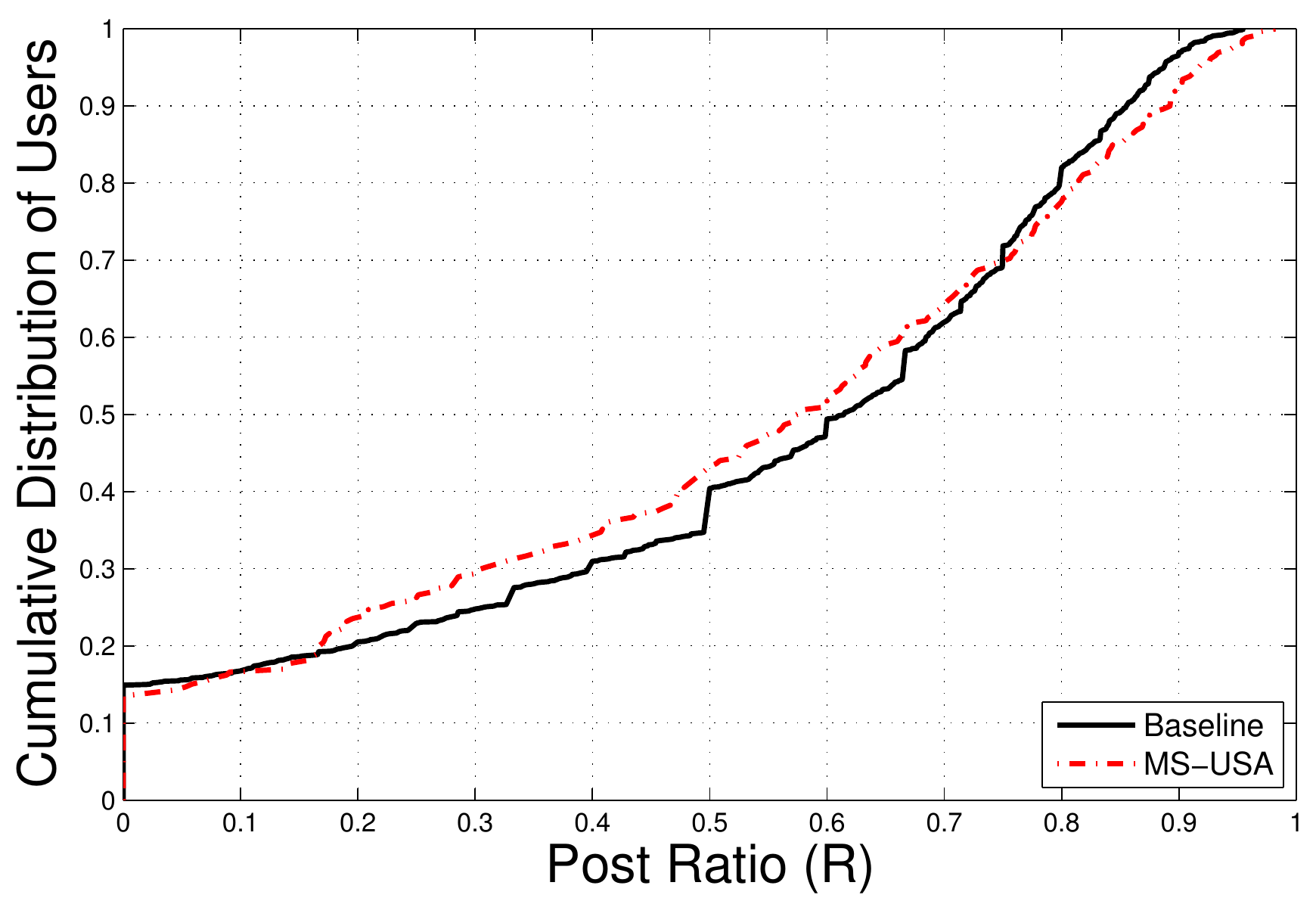}
		\label{fig:ms_us_n_sf}
	}
\vspace{-0.15cm}	
	\caption{Distributions of the ratio of English posts to non-English posts.}%
	\label{fig:lang_ratio}
\end{figure*}

\subsection{Analysis of Lexical Features}
\label{subsec:lexicalanalysis}
We now look at features that relate to the content of timeline posts, similar lexical features could be extracted for other non-English languages. We acknowledge that the extraction of lexical features of a non-English language is a challenging task and the extraction models might be prone to errors. We constrain our analysis to only English language and argue that lexical features extractions and analysis could be extended for other non-English Language such as Chinese~\cite{Zhang:2003:CLA,Zhang:2003:HCL:1119250.1119280}, French~\cite{Silberztein1989}, and Arabic~\cite{farghaly2009arabic}, and Hindi/Urdu~\cite{tiwary2008natural}. We refer the reader to~\cite{silberztein1997lexical} for more details about lexical features used in this paper. 

We have also considered user timelines as the collection of posts and the corresponding comments on each post (i.e., all textual content) and build a corpus of words extracted from the timelines by applying the term frequency-inverse document frequency (TF-IDF) statistical tool~\cite{Salton:tfidf}. However, the overall performance of this ``bag-of-words'' approach was poor, which can be explained with the short nature of the posts. Indeed, \cite{Hogenboom:2015} has shown that the word frequency approach to analyze short text on social media and blogs does not perform well. Thus, in our work, we disregard simple TF-IDF based analysis of user timelines and identify other lexical features.

\descr{Language.} Next, we analyze the ratio of posts in English, i.e., for every post we filter out all non-English ones using a standard language detection library.\footnote{\url{https://python.org/pypi/langdetect} [Accessed on July 18$^{th}$, 2016].} For each user, we count the number of English-language posts and calculate its ratio with respect to the total number of posts. \figurename~\ref{fig:lang_ratio} shows that the baseline users and like farm users in USA (i.e., MS-USA, BL-USA, and AL-USA) mostly post in English, while users of worldwide campaigns (MS-ALL, BL-ALL, AL-ALL) have significantly fewer posts in English. For example, the median ratio of English posts for AL-ALL campaign is around 10\% and that for SF-ALL around 15\%. We acknowledge that our analysis is limited to English-only content and may be statistically biased toward non-native English speakers i.e., non-USA campaign users. While our analysis could be extended to other languages, we argue that English-based lexical analysis provides sufficient differences across different categories of users. Thus, developing algorithms for language detection and processing on non-English posts is out of the scope of this paper.

\descr{Readability.} We further analyze posts for grammatical and semantic correctness. We parse each post to extract the number of words, sentences, punctuation, non-letters (e.g., emoticons), and measure the lexical richness, as well as the Automated Readability Index (ARI)~\cite{smith:ari} and Flesch score \cite{flesch48readability}. Lexical richness, defined as the ratio of number of unique words to total number of words, reveals noticeable repetitions of distinct words, while the ARI, computed as 4.71 $\times$ average word length) + (0.5 $\times$ average sentence length) - 21.43, estimates the comprehensibility of a text corpus. Table \ref{tab:lexicalfeatures} shows a summary of the results. In comparison to like farm users, baseline users post text with higher lexical richness (70\% vs. 55\%), ARI (20 vs. 15), and Flesch score (55 vs. 48), thus suggesting that normal users use a richer vocabulary and that their posts have higher readability.

\begin{table*}[!t]
\begin{center}
\small
\tabcolsep=0.12cm
\begin{tabular}{lrrrrrrrrrrrr}
\toprule
{\bf Campaign}&  	\bf Avg 	& \bf Avg & \bf Avg & {\bf Avg Sent} & {\bf Avg Word}  & {\bf Richness} & {\bf ARI}  & {\bf Flesch}\\
 &  {\bf Chars}	&  {\bf Words}	& {\bf Sents} &  {\bf Length}  & {\bf Length} & & & {\bf Score}\\
\midrule
Baseline	&	4,477	&	780	&	67	&	6.9	&	17.6	&	0.70	&	20.2	&	55.1	\\
\midrule

BL-USA	&	7,356	&	1,330	&	63	&	5.7	&	22.8	&	0.58	&	16.9	&	51.5	\\
AL-ALL	&	2,835	&	464	&	32	&	6.2	&	13.9	&	0.59	&	14.8	&	43.6	\\
AL-USA	&	2,475	&	394	&	33	&	6.2	&	12.7	&	0.49	&	14.1	&	54.0	\\
SF-ALL	&	1,438	&	227	&	19	&	6.3	&	11.7	&	0.58	&	14.1	&	45.2	\\
SF-USA	&	1,637	&	259	&	22	&	6.3	&	12.0	&	0.55	&	14.4	&	45.6	\\
MS-USA	&	6,227	&	1,047	&	66	&	6.1	&	17.8	&	0.53	&	16.2	&	50.1	\\
\bottomrule
\end{tabular}
\vspace{-0.15cm}
\caption{Lexical analysis of timeline posts.}
\label{tab:lexicalfeatures}		
\end{center}
\end{table*}

\subsection{Remarks}
Our analysis of user timelines highlights several differences in both lexical and non-lexical features of normal and like farm users. In particular, we find that posts made by like farm accounts have 43\% fewer words, a more limited vocabulary, and lower readability than normal users' posts. Moreover, %
like farm users generate significantly more comments and likes and a large fraction of their posts consists of non-original and often redundant ``shared activity''. 

In the next section, we will use these timelines features to automatically detect like farm users using a machine learning classifier.

\section{Timeline-based Detection of Like Farms}
\label{sec:detection}
Aiming to automatically distinguish like farm users from normal (baseline) users, we use a supervised two-class SVM classifier~\cite{Muller01anintroduction}, implemented using \emph{scikit-learn}~\cite{sklearn_api} (an open source machine learning library for Python). We later compare this classifier with other well-known supervised classifiers such as Decision Tree~\cite{dtree}, AdaBoost~\cite{adaboost}, kNN~\cite{knn}, Random Forest~\cite{Breiman:rf}, and Na\"ive Bayes~\cite{zhang2004optimality} and confirm that the two-class SVM is the most effective in detecting like farms users.

We extract four non-lexical features and twelve distinct lexical features from the timelines of baseline and like farm users, as explained in Section~\ref{sec:characterizing}, using the datasets presented in Section~\ref{sec:data-new}. The non-lexical features are the average number of words, comments, likes per post, and re-shares. The lexical features include: the number of characters, words, and sentences; the average word length, sentence length, and number of upper case letters; the average percentage of punctuation, numbers, and non-letter characters; richness, ARI, and Flesch Score.

We form two classes by labeling like farm and baseline users' lexical and non-lexical features as positives and negatives, respectively. We use 80\% and 20\% of the features to build the training and testing sets, respectively. Appropriate values for parameters $\gamma$ (\emph{radial basis function kernel} parameter~\cite{Scholkopf:2001:ESH:1119748.1119749}) and $\upsilon$ (SVM parameter) are set empirically by performing a greedy grid search on ranges $2^{-10} \leq \gamma \leq 2^{0}$ and $2^{-10} \leq \upsilon \leq 2^{0}$, respectively, on each training group.

\begin{table*}[t]
\tabcolsep=0.11cm
  \begin{center}
	\small
\resizebox{0.7\linewidth}{!}{
    \begin{tabular}{crrrrrrrrrrr}
      \toprule
			\bf Campaign	&	\bf Total	&	\bf  Training	&	\bf Testing 	&	\bf TP	&	\bf FP	&	\bf TN	&	\bf FN	&	\bf Precision	&	\bf Recall	&	\bf Accuracy	 &	\bf F1-\\
& \bf Users & \bf Set & \bf Set & & & & & & & & \bf Score \\
\midrule

BL-USA	&	583	&	466	&	117	&	37	&	80	&	270	&	12	&	76\%	&	32\%	&	77\%	&	45\%	\\
AL-ALL	&	707	&	566	&	141	&	132	&	9	&	278	&	4	&	96\%	&	94\%	&	97\%	&	95\%	\\
AL-USA	&	827	&	662	&	164	&	113	&	51	&	278	&	4	&	97\%	&	69\%	&	88\%	&	81\%	\\
SF-ALL	&	870	&	696	&	174	&	139	&	35	&	273	&	9	&	94\%	&	80\%	&	90\%	&	86\%	\\
SF-USA	&	653	&	522	&	131	&	110	&	21	&	277	&	5	&	96\%	&	84\%	&	94\%	&	90\%	\\
MS-USA	&	259	&	207	&	52	&	39	&	13	&	280	&	2	&	95\%	&	75\%	&	96\%	&	84\%	\\
\bottomrule
    \end{tabular}
    }
    \caption{Effectiveness of non-lexical features (+SVM) in detecting like farm users.}
    \label{tab:svm_on_non_lexical_features}
      \end{center}
\end{table*}

\descr{Non-Lexical Features.}
Table~\ref{tab:svm_on_non_lexical_features} reports on the accuracy of our classifier with non-lexical features, i.e., users interactions with posts as described in Section~\ref{sec:charNL}. Note that for each campaign, we train the classifier with 80\% of the non-lexical features from baseline and campaign training sets derived from the campaign users timelines. The poor classification performance for the stealthiest like farm (BL-USA) suggests that non-lexical features alone are not sufficient to accurately detect like farm users.

\begin{table*}[t]
\tabcolsep=0.11cm
  \begin{center}
   \small
\resizebox{0.7\linewidth}{!}{
    \begin{tabular}{crrrrrrrrrrr}
      \toprule
			\bf Campaign	&	\bf Total	&	\bf  Training	&	\bf Testing 	&	\bf TP	&	\bf FP	&	\bf TN	&	\bf FN	&	\bf Precision	&	\bf Recall	&	\bf Accuracy	 &	\bf F1-	\\
& \bf Users & \bf Set & \bf Set & & & & & & & & \bf Score \\
\midrule
BL-USA	&	564	&	451	&	113	&	113	&	0	&	240	&	0	&	100\%	&	100\%	&	100\%	&	100\%	\\

AL-ALL	&	675	&	540	&	135	&	133	&	2	&	238	&	2	&	99\%	&	99\%	&	99\%	&	99\%	\\
AL-USA	&	570	&	456	&	114	&	113	&	1	&	239	&	1	&	99\%	&	99\%	&	99\%	&	99\%	\\
SF-ALL	&	761	&	609	&	152	&	151	&	1	&	238	&	2	&	99\%	&	99\%	&	99\%	&	99\%	\\
SF-USA	&	570	&	456	&	114	&	113	&	1	&	225	&	15	&	99\%	&	87\%	&	95\%	&	92\%	\\
MS-USA	&	224	&	179	&	45	&	45	&	0	&	240	&	0	&	100\%	&	100\%	&	100\%	&	100\%	\\
\bottomrule
    \end{tabular}
   }
\vspace{-0.15cm}   
      \caption{Effectiveness of lexical features (+SVM) in detecting like farm users.}
      \label{tab:svm_on_lexical_features}
  \end{center}
\end{table*}

\descr{Lexical Features.}
Next, we evaluate the accuracy of our classifier with lexical features, reported in Table~\ref{tab:svm_on_lexical_features}. We filter out all users with no English-language posts (i.e., with the ratio of English posts to non-English posts, R=0, see \figurename~\ref{fig:lang_ratio}). Again, we train the classifier with 80\% lexical features from baseline and like farm training sets. We observe that our classifier achieves very high precision and recall for MS-USA, BL-USA, and AL-USA. Although the accuracy decreases by approximately 8\% for SF-USA, the overall performance suggests that lexical features are useful in automatically detecting like farm users.

\begin{table*}[!t]
\tabcolsep=0.11cm
  \begin{center}
\resizebox{0.7\linewidth}{!}{
    \begin{tabular}{crrrrrrrrrrr}
      \toprule
			\bf Campaign	&	\bf Total	&	\bf  Training	&	\bf Testing 	&	\bf TP	&	\bf FP	&	\bf TN	&	\bf FN	&	\bf Precision	&	\bf Recall	&	\bf Accuracy	 &	\bf F1-\\
& \bf Users & \bf Set & \bf Set & & & & & & & & \bf Score \\
\midrule
BL-USA	&	583	&	466	&	117	&	116	&	1	&	278	&	4	&	99\%	&	97\%	&	99\%	&	98\%	\\
AL-ALL	&	707	&	566	&	141	&	140	&	1	&	278	&	4	&	99\%	&	97\%	&	99\%	&	98\%	\\
AL-USA	&	827	&	662	&	164	&	164	&	0	&	275	&	7	&	100\%	&	96\%	&	98\%	&	97\%	\\
SF-ALL	&	870	&	696	&	174	&	172	&	2	&	271	&	11	&	99\%	&	94\%	&	97\%	&	96\%	\\
SF-USA	&	653	&	522	&	131	&	130	&	1	&	273	&	9	&	99\%	&	93\%	&	98\%	&	96\%	\\
MS-USA	&	259	&	207	&	52	&	52	&	0	&	280	&	2	&	100\%	&	96\%	&	99\%	&	98\%	\\
\bottomrule
    \end{tabular}
}
\vspace{-0.15cm}
    \caption{Effectiveness of both lexical and non-lexical features (+SVM) in detecting like farm users.}
    \label{tab:svm_on_all_features}
      \end{center}
\end{table*}

\descr{Combining Lexical and Non-Lexical Features.}
While building a classifier based on lexical features performs very well in detecting fake accounts, we acknowledge that lexical features may be affected by geographical location especially if one set of users who write in English are native speakers while the other set is not.
Therefore, we further combine both lexical and non-lexical features to build a more robust classifier.
We also note that approximately 3\% to 22\% of like farm users and 14\% of baseline users do not have English language posts and are not considered in the lexical features based classification.
To include these users in our classification, for each like farm and baseline, we set their lexical features to zeros and aggregate the lexical features with non-lexical features and evaluate our classifier with the same classification methodology as detailed above.
Results are summarized in Table~\ref{tab:svm_on_all_features}, which shows high accuracy for all like farms (F1-Score $\geq$ 96\%), thus confirming the effectiveness of our timeline-based features in detecting like farm users.

\begin{table*}[!t]
\tabcolsep=0.11cm
  \begin{center}
  		  \resizebox{0.7\textwidth}{!}{%
    \begin{tabular}{ccccccc}
      \toprule
{\bf Campaign}	 & \bf SVM& \bf Decision Tree	& \bf AdaBoost & \bf	kNN	& \bf Random Forest & \bf Na\"ive Bayes\\
\midrule
BL-USA	&	98\%	&	96\%	&	96\%	&	91\%	&	88\%	&	53\%	\\
AL-ALL	&	98\%	&	84\%	&	95\%	&	86\%	&	84\%	&	75\%	\\
AL-USA	&	97\%	&	88\%	&	90\%	&	91\%	&	86\%	&	81\%	\\
SF-ALL	&	96\%	&	90\%	&	94\%	&	89\%	&	87\%	&	67\%	\\
SF-USA	&	96\%	&	83\%	&	92\%	&	79\%	&	78\%	&	61\%	\\
MS-USA	&	98\%	&	90\%	&	89\%	&	89\%	&	87\%	&	74\%	\\
\bottomrule
    \end{tabular}
}
\vspace{-0.15cm}
    \caption{F1-Score obtained with different classification methods, using both lexical and non-lexical features, in detecting like farm users.}
    \label{tab:classifiers_on_all_features}
      \end{center}
\end{table*}

\descr{Comparison With Other Machine Learning Classifiers.}
In order to generalize our approach, we have also used other machine learning classification algorithms, i.e., Decision Tree, AdaBoost, kNN, Random Forest, and Na\"ive Bayes. The training and testing of all these classifiers follow the same set-up as the SVM approach. We again use 80\% and 20\% of the combined lexical and non-lexical features to build the training and testing sets, respectively. We summarize the performance of the classifiers in Table~\ref{tab:classifiers_on_all_features}.  Our results show that the SVM classifier achieves the highest F1-Scores across the board. Due to overfitting on our dataset, Random Forest and Na\"ive Bayes show poor results and require mechanism such as pruning, detailed analysis of parameters, as well as selection of the optimal set of prominent features to improve classification performance~\cite{kohavi1995feature,Breiman:rf}.

\begin{figure*}[!t]
	\centering
\subfigure[]{\includegraphics[width=0.82\columnwidth]{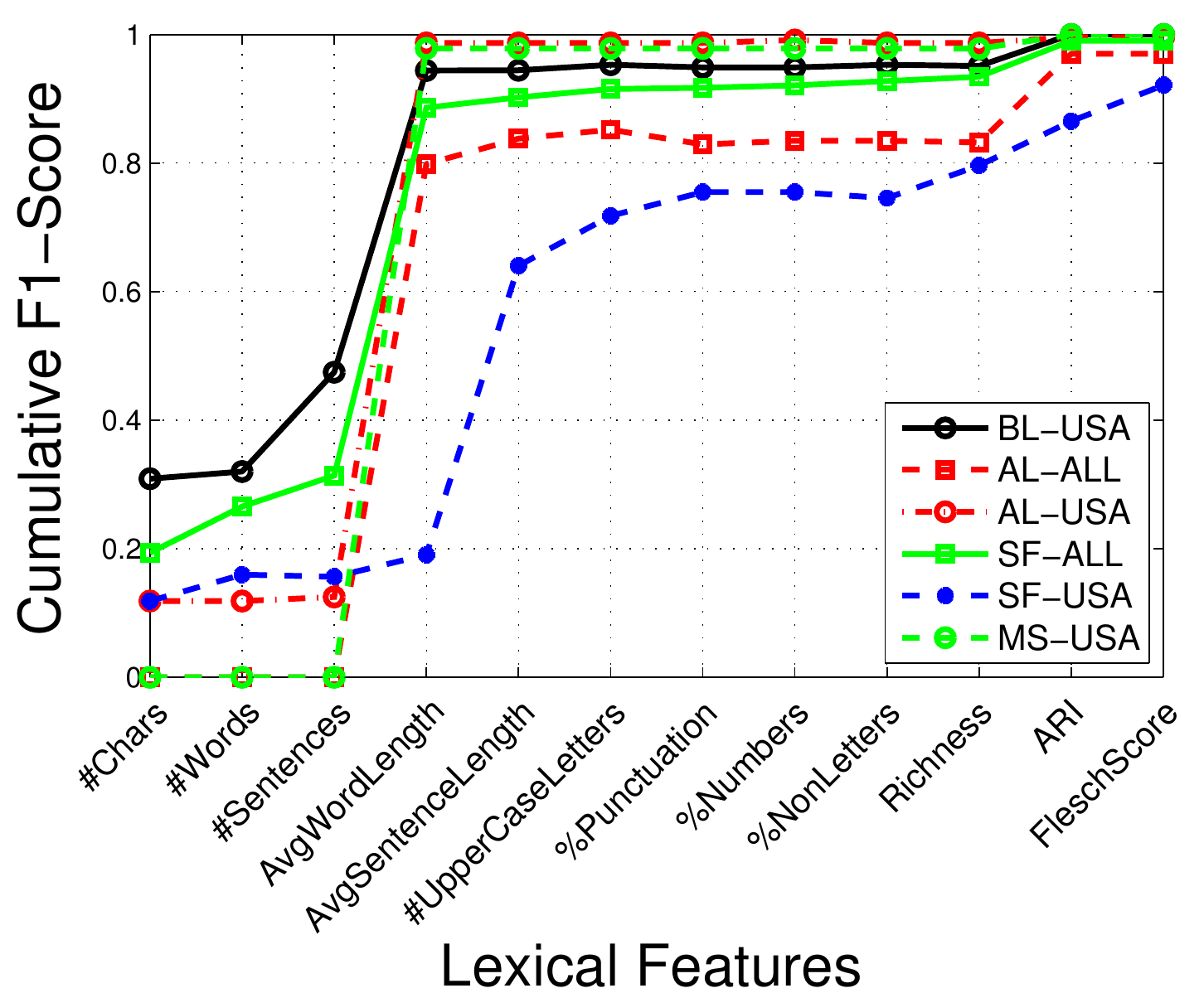}\label{fig:lexical_fa}}
\subfigure[]{\includegraphics[width=0.84\columnwidth]{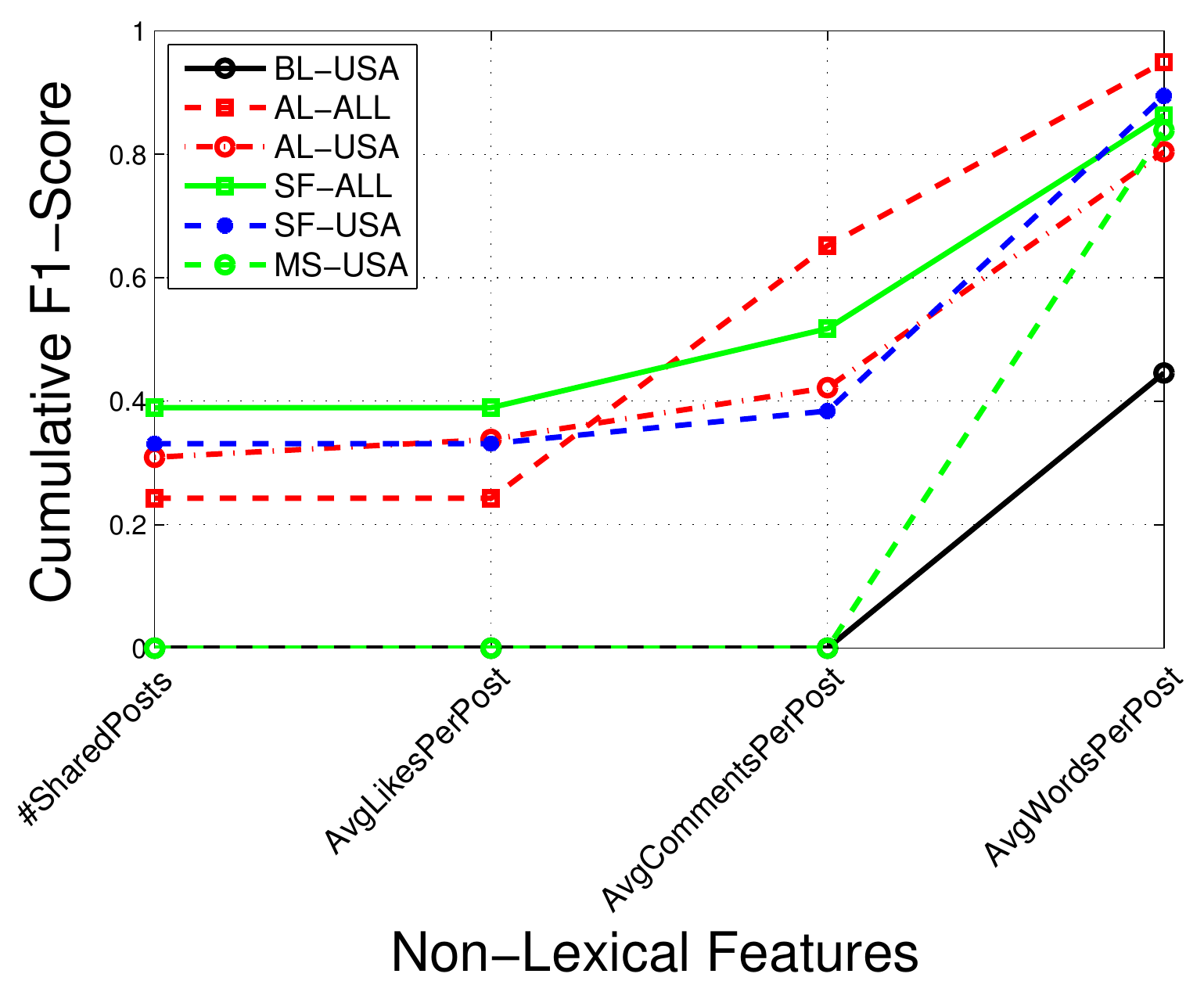}\label{fig:nonlexical_fa}}\\
\subfigure[]{\includegraphics[width=0.82\columnwidth]{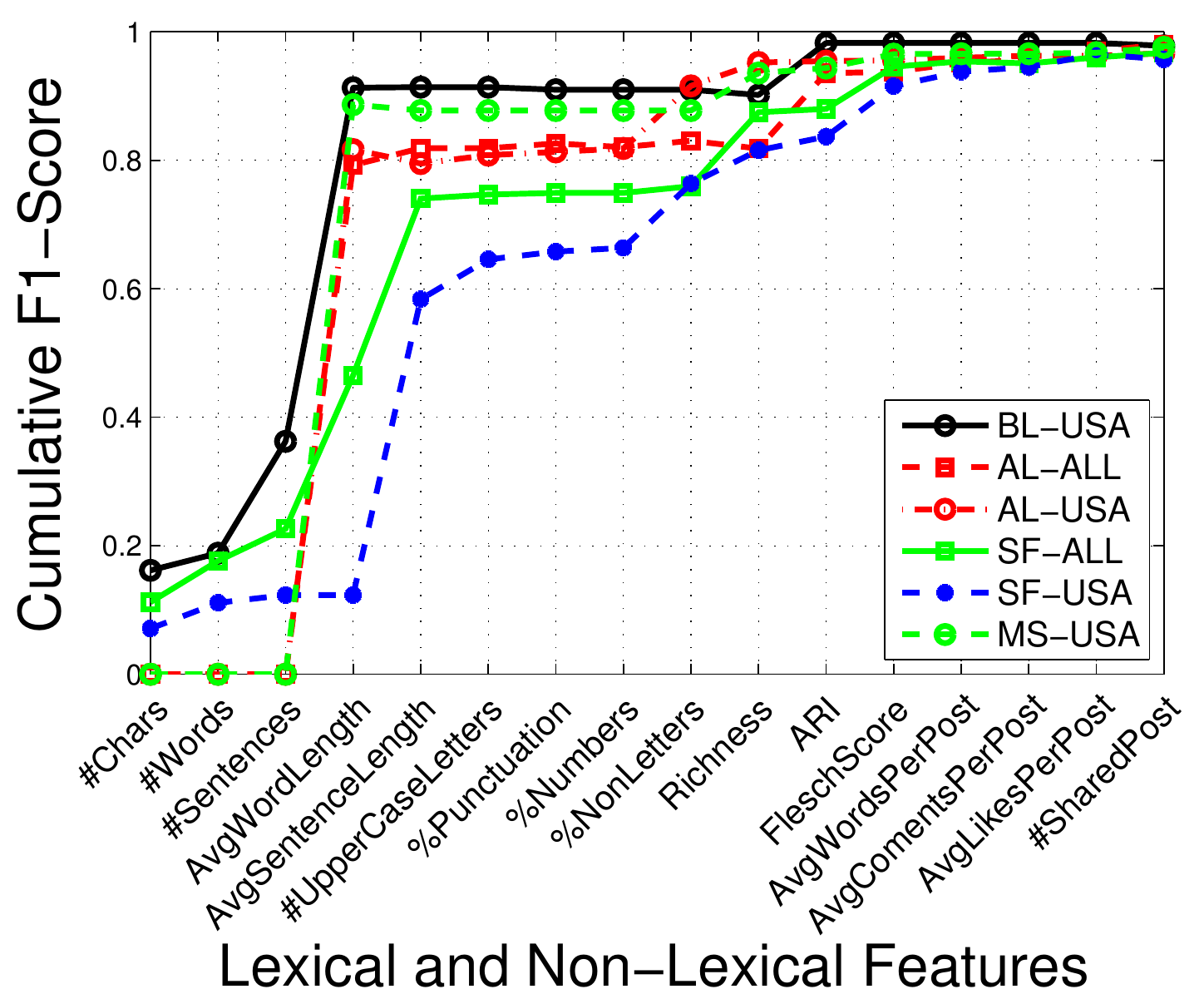}\label{fig:combined_fa}}
\vspace{-0.2cm}
\caption{Cumulative F1-Score for all lexical and non-lexical features measured. The X-axis shows the incremental inclusion of features in both training and testing of SVM. Details of the classification performance for all features are listed in Table~\ref{tab:svm_on_all_features}. }%
\label{fig:factoranalysis}
\vspace{0.3cm}
\end{figure*}

\descr{Analysis.}
We now analyze in more details the classification performance (in terms of F1-Score) to identify the most distinctive features. Specifically, we incrementally add lexical and non-lexical features to train and test our classifier for all campaigns. We observe that the average word length (cf. \figurename~\ref{fig:lexical_fa}) and average number of words per post (cf. \figurename~\ref{fig:nonlexical_fa}) provide the most improvement in the F1-Score for all campaigns. This finding suggests that like farm users use shorter words and fewer number of words in their timeline posts as compared to baseline users. While these features provide the largest improvement in detecting a like farm account, an attempt to circumvent detection by increasing the word length or number of words per post will also effect the ARI, Flesch score, and richness. That is, increasing word length and number of words on posts in a way that is not readable nor understandable, will not improve the overall outlook of the account to appear real. Therefore, combining several features increases the workload required to appear real on like farm accounts. %
The overall classification accuracy with both lexical and non-lexical features is reported in \figurename~\ref{fig:combined_fa}.

\begin{figure*}[!t]
	\centering
	\subfigure[]{\includegraphics[width=0.8\columnwidth]{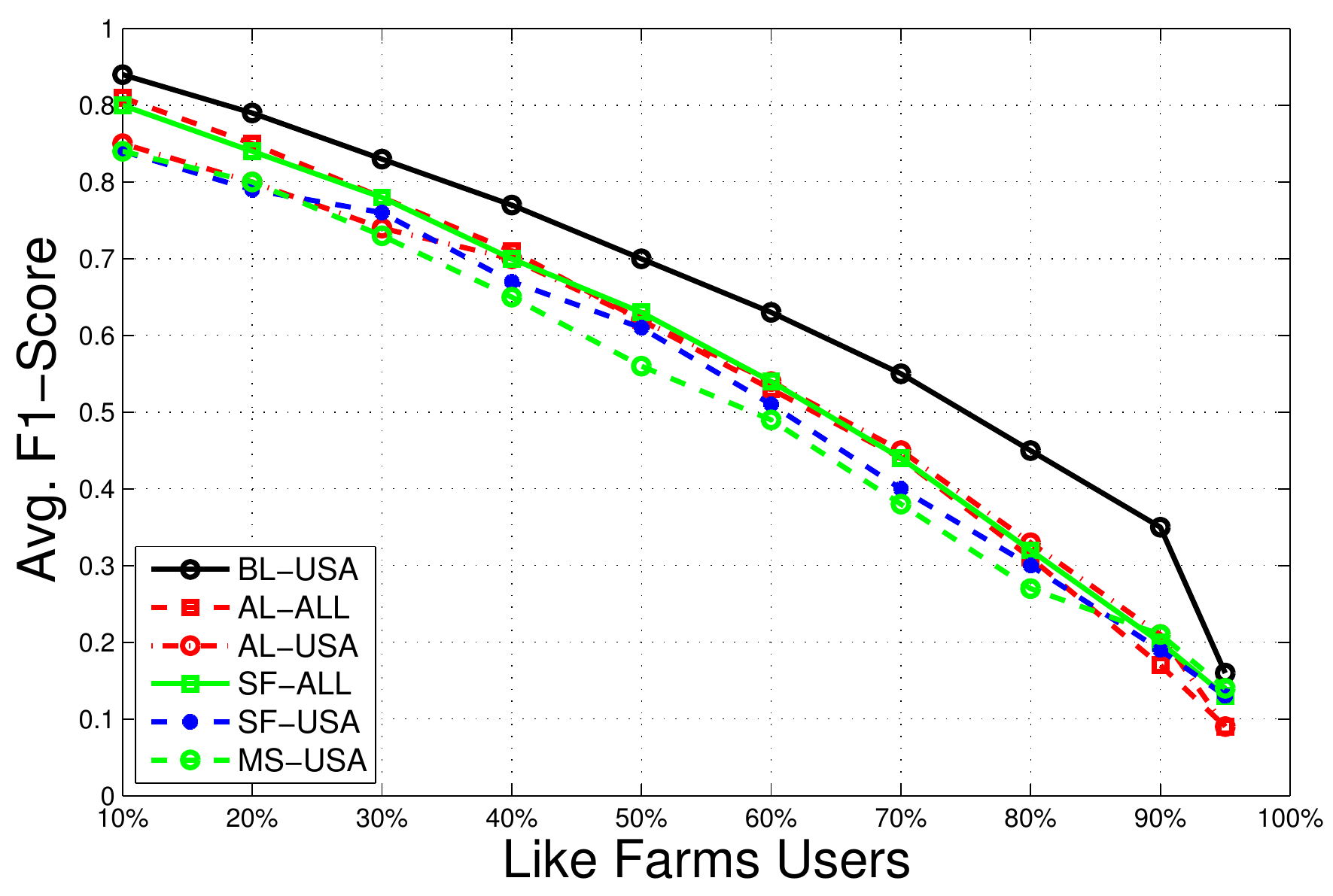}\label{fig:f1_score_lf_mimicking}}
\subfigure[]{\includegraphics[width=0.8\columnwidth]{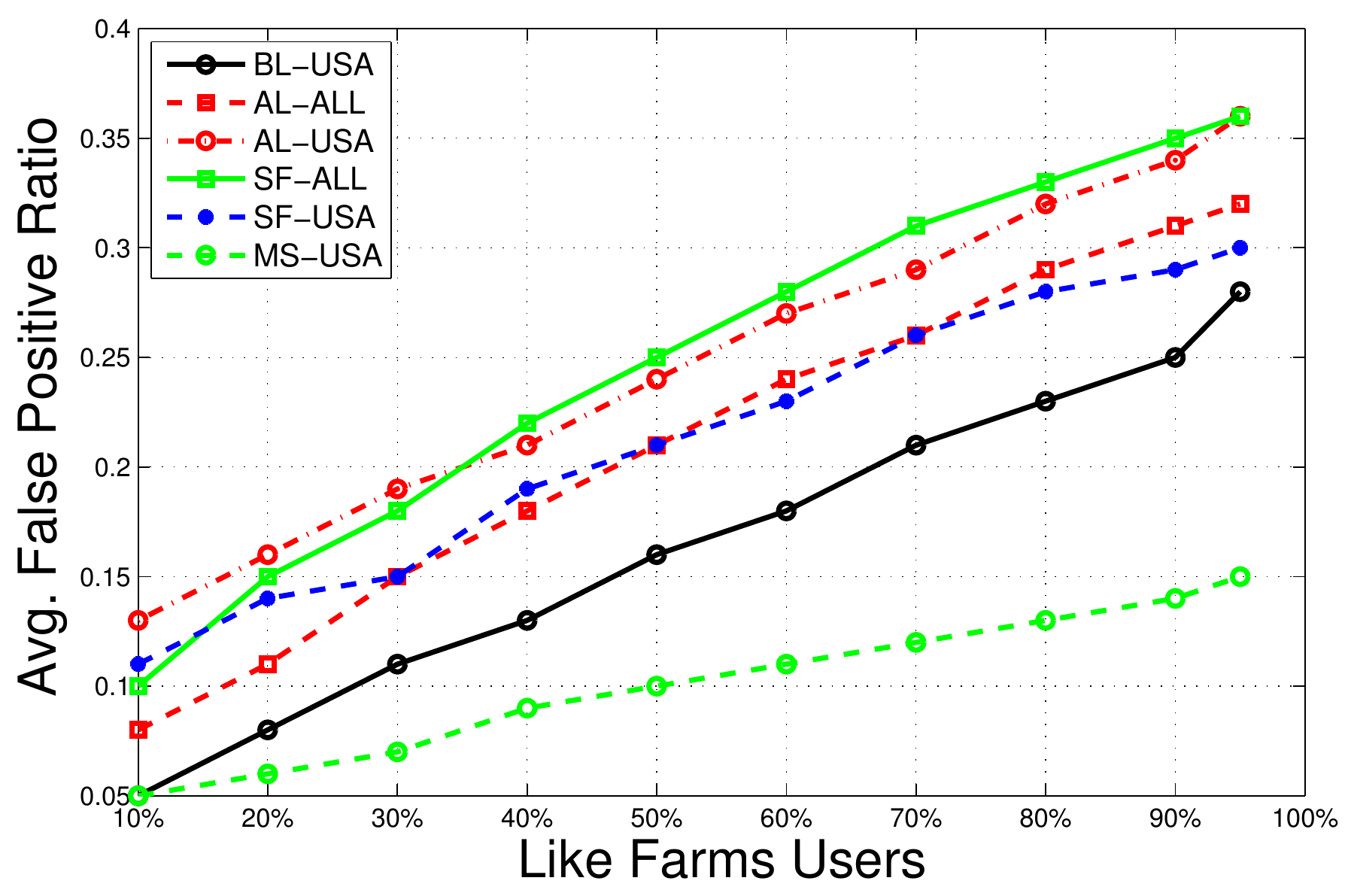}\label{fig:f1_score_fpr_mimicking}}
\vspace{-0.15cm}
\caption{Average F1-Score and false positive ratio measured when fractions of like farms users mimic all lexical and non-lexical features (+SVM). The X-axis shows the percentage of like farms users that are mimicking baseline users.}\label{fig:robustnessanalysis}
\end{figure*}

\descr{Robustness Of Our Approach.} The like farms users may evade our detection system by mimicking the behavior of real users. To test the effectiveness of our features and classifiers, we assume two worst case attacking scenarios: \textit{(i)} fractions of like farms users mimic all features of baseline users; and \textit{(ii)} all like farm users mimic sets of baseline users' features
We simulate the first scenario by assuming that sets of like farm users randomly select baseline users and aggressively replace the values of all their features with that of the selected baseline users. We use the aforementioned settings of the best of our classifiers, SVM, and run the experiments for each like farm 10 times. Figure~\ref{fig:robustnessanalysis} shows the effect on F1-Score of our classifier when fractions of like farm users aggressively mimic all the lexical and non-lexical features of baseline users. When 30\% of like farms users coordinate and mimic all features of baseline users, we observe that our classifier achieves at least 73\% F1-Score and at most 17\% false positive ratio, decreasing 26\% F1-Score compared to our approach (cf. Table~\ref{tab:svm_on_all_features}).
For the latter case, we assume that all like farms users coordinate and select sets of features %
from randomly selected baseline users that they copy or mimic. We use identical configuration of our SVM classifier, and conduct experiments for each like farm 10 times. Table~\ref{tab:features_robustness} summarizes the results of our experiments. With this attack strategy, we observe that when only one feature is mimicked, %
the F1-Score of our approach (cf. Table~\ref{tab:svm_on_all_features}) decreases by between 1\% to 8\%. The F1-Score of our classifier decreases by between 26\% to 56\% when the like farm users target sets of 8 features including prominent ones (cf. Figure~\ref{fig:factoranalysis}).
Note that any feature used to identify fake like farms behavior can be either circumvented or manipulated by the like farms users by behaving more like real users. We believe that this is a typical arm-race that eventually raise the bar for the like farms -- the more effort they need to invest in appearing as real users, the lower their incentive is to do this.

\begin{table*}[!t]
\tabcolsep=0.11cm
  \begin{center}
  		  \resizebox{0.55\textwidth}{!}{%
    \begin{tabular}{cccccc}
      \toprule
      \multicolumn{6}{c}{\bf $\Delta$ F1-Score} \\
\cmidrule(r){2-6}
{\bf Campaign}	 & \bf 1-Feature & \bf 2-Features	& \bf 3-Features & \bf	4-Features	& \bf 8-Features \\
\midrule
BL-USA	&	1\%	&	2\%	&	3\%	&	5\%	&	42\%\\
AL-ALL	&	2\%	&	3\%	&	4\%	&	5\%	&	47\%	\\
AL-USA	&	2\%	&	4\%	&	5\%	&	10\%	&	20\%	\\
SF-ALL	&	3\%	&	4\%	&	6\%	&	6\%	&	56\%\\
SF-USA	&	8\%	&	9\%	&	11\%	&	13\%	&	55\%	\\
MS-USA	&	5\%	&	6\%	&	6\%	&	7\%	&	26\%	\\
\bottomrule
    \end{tabular}
}
\vspace{-0.15cm}
    \caption{The difference in F1-Score obtained when all like farm users coordinate and mimic sets of lexical and non-lexical features of baseline users. F1-Score in Table~\ref{tab:svm_on_all_features} is used as a reference to compute the $\Delta$ in F1-Score.}
    \label{tab:features_robustness}
      \end{center}
\end{table*}

\descr{Remarks.}
Our results demonstrate that it is possible to accurately detect like farm users from both sophisticated and na\"ive farms by incorporating additional account information -- specifically, timeline activities. 
The low false positive ratio ($\ll$1\%, cf. Table~\ref{tab:svm_on_all_features}) highlights the effectiveness of our approach as well as the limitations of prior graph co-clustering algorithms in detecting like farms users (cf. Section~\ref{sec:coclustering}). Unfortunately, we do not have access to a larger dataset to measure and discuss the effects on false positive ratio of our approach. We believe then that without an evaluation of our approach at a larger scale, further discussion would be speculative so we refrained from further interpretation of those results.
We also argue that the use of a variety of lexical and non-lexical features will make it difficult for like farm operators to circumvent detection. Like farms typically rely on pre-defined lists of comments, resulting in word repetition and lower lexical richness. As a result, we argue that, should our proposed techniques be deployed by Facebook, it will be challenging, as well as costly, for fraudsters to modify their behavior and evade detection, since this would require instructing automated scripts and/or cheap human labor to match the diversity and richness of real users' timeline posts.

\section{Related Work}
\label{sec:related}
Prior work has focused quite extensively on the analysis and the detection of sybil and/or fake accounts in online social networks by relying on tightly-knit community structures~\cite{yu06sybilguard,danezis09SybilInfer,yang11socialnetworksybils,cao12fakeosn,yang12spammersocialnetwork,boshmaf15integro}. %
 By contrast, we work to detect accounts that are employed by like farms to boost the number of Facebook page likes, whether they are operated by a bot or a human. We highlight several characteristics about the social structure and activity of fake profiles attracted by the honeypot pages, e.g., their interconnected nature or the activity bursts. In fact, our analysis does not only confirm a few insights used by sybil detection algorithms but also reveals new patterns that could complement them.
Fraud and fake activities are not restricted to social network, but widespread also on other platforms, such as online gaming. In this context,~\cite{lee2016you}
rely on self-similarity to effectively measure
the frequency of repeated activities per player over time, and use it to identify bots.
Also,~\cite{kwon2017crime}  analyze the characteristics of the ecosystem of multiplayer online role-playing games (MMORPGs), and devise a method for detecting gold farming groups (GFGs), based on graph techniques.

Prior work on reputation manipulation on social networks include a few {\em passive} measurement studies have also focused on characterizing fake user accounts and their activity.
Nazir \etal~\cite{nazir10facebookphantomprofiles} studied phantom profiles in Facebook gaming applications, while
Thomas \etal~\cite{thomas11suspendedaccounts} analyzed over 1.1 million accounts suspended by Twitter.
Gao \etal~\cite{gao10socialspamcampaigns} studied spam campaigns on Facebook originating from approximately 57,000 user accounts.
Yang \etal~\cite{yang12spammersocialnetwork} performed an empirical analysis of social relationships between spam accounts on Twitter, and Dave \etal~\cite{dave2012measuring} proposed a methodology to measure and fingerprint click-spam in ad networks.
Our work differs from these studies as they all conducted passive measurements, whereas we rely on the deployment of several honeypot pages and (paid) campaigns to actively engage with fake profiles. Lee \etal~\cite{lee10socialspamhoneypots} and Stringhini \etal~\cite{stringhini10spammerssocialnetworks} created honeypot profiles in Facebook, MySpace, and Twitter to detect spammers while we use accounts attracted by our honeypot Facebook pages that actively engage like farms. Unlike~\cite{lee10socialspamhoneypots,stringhini10spammerssocialnetworks}, we
leverage timeline-based features for the detection of fake accounts. Our work also differs from theirs in that
(1) their honeypot profiles were designed to look legitimate, while our honeypot pages explicitly indicated they were not ``real'' (to deflect real profiles), and (2) our honeypot pages {\em actively} attracted fake profiles by means of paid campaigns, as opposed to passive honeypot profiles.

Thomas \etal \cite{thomas13traffickingfraudtwitteraccounts} analyzed trafficking of fake accounts in Twitter. They bought fake profiles from 27 merchants and developed a classifier to detect these fake accounts. In a similar study, Stringhini \etal~\cite{stringhini12twitterfollowermarketWOSN,stringhini13twitterfollower} analyzed the market of {\em Twitter followers}, which, akin to Facebook like farms, provide Twitter followers for sale.
Note that Twitter follower markets differ from Facebook like farms as Twitter entails a {\em follower-followee} relationship among users, while Facebook friendships imply a bidirectional relationships. Also, there is no equivalent of liking a Facebook page in the Twitter ecosystem.

Wang et al.~\cite{wang14adversarialdetection} studied human involvement in Weibo's reputation manipulation services,
showing that simple evasion attacks (e.g., workers modifying their behavior) as well as poisoning attacks (e.g., administrators tampering with the training set) can severely affect the effectiveness of machine learning algorithms to detect malicious crowd-sourcing workers. Song \etal~\cite{song2015crowdtarget} also looked at {\em crowdturfing} services that manipulate account popularity on Twitter through artificial retweet and developed ``CrowdTarget" to detect such tweets.
Partially informed by these studies, we do not only cluster like activity performed by users but also build on lexical and non-lexical features.
Specific to Facebook fraud is CopyCatch~\cite{beutel2013copycatch},
a technique deployed by Facebook to detect fraudulent accounts by identifying groups of connected users liking a set of pages within a short time frame.
SynchroTrap~\cite{cao14synchrotrap} extended CopyCatch by clustering accounts that perform similar, possibly malicious, synchronized actions, using tunable parameters such as time-window and similarity thresholds in order to improve detection accuracy.
However, as discussed earlier, while some farms seem to be operated by bots (producing large bursts of likes and having limited numbers of friends) that do not really try to hide their activities, other {\em stealthier} farms exhibit behavior that may be challenging to detect with tools like CopyCatch and SynchroTrap.
In fact, our evaluation of graph co-clustering techniques shows that these farms successfully evade detection by avoiding lockstep behavior and liking sets of seemingly random pages.
As a result, we use timeline features, relying on both lexical and non-lexical features to build a classifier that detects stealthy like farm users with high accuracy.
Finally, we highlight that our work can complement other methods used in prior work to detect fake and compromised accounts, such as using unsupervised anomaly detection techniques~\cite{viswanath14tanomaloussocialnetwork}, temporal features~\cite{jiang14catchsyn,jiang14strangebehaviorosocial}, IP addresses~\cite{stringhini2015evilcohort},
as well as generic supervised learning~\cite{badri2016uncovering}.

\descr{Remarks on ``New Material''.} Compared to our preliminary results (published in~\cite{decristofaro14facebooklikefarms} and reported in Section~\ref{sec:imc}), this article clearly introduces significant additional new material. Specifically: (i) we introduce an empirical evaluation demonstrating that temporal and social graph analysis can only be used to detect naive farms (Section~\ref{sec:coclustering}), and
(ii) we present a novel timeline-based classifier geared to detect accounts from stealthy like farms with a remarkably high degree of accuracy (Sections~\ref{sec:characterizing} and~\ref{sec:detection}).

\section{Conclusion}
\label{sec:conclusion}
Minimizing fraud in online social networks is crucial for maintaining the confidence and trust of the user base and investors. In this paper, we presented the results of a measurement study of Facebook like farms -- i.e., paid services artificially boosting the number of likes on a Facebook page -- aiming to identify characteristics and accurately detect the accounts used by them.
We crawled profile information, liking patterns, and timeline activities from like farms accounts. Our demographic, temporal, and social graph analysis highlighted similar patterns between accounts across different like farms and revealed two main modi operandi: some farms seem to be operated by bots and do not really try to hide the nature of their operations, while others follow a stealthier approach, mimicking regular users' behavior.
We then evaluated the effectiveness of existing graph based fraud detection algorithms, such as CopyCatch~\cite{beutel2013copycatch} and SynchroTrap~\cite{cao14synchrotrap}, and demonstrated that sophisticated like farms can successfully evade detection.

Next, aiming to address their shortcomings, we  focused on incorporating additional profile information from accounts' timelines in order to train machine learning classifiers geared to distinguish between like farm users from normal ones.
We extracted lexical and non-lexical features from user timelines, finding that posts by like farm accounts have 43\% fewer words, a more limited vocabulary, and lower readability than normal users' posts. Moreover, like farm posts generated significantly more comments and likes, and a large fraction of their posts consists of non original and often redundant ``shared activity'' (i.e., repeatedly sharing posts made by other users, articles, videos, and external URLs). By leveraging both lexical and non-lexical features, we experimented with several machine learning classifiers, with the best of our classifiers (SVM) achieving as high as 100\% precision and 97\% of recall, and at least 99\% and 93\% respectively across all campaigns -- significantly higher than graph co-clustering techniques.

In theory, fraudsters could try to modify their behavior in order to evade our proposed timeline-based detection. However, like farms either heavily automate mechanisms or rely on manual input of cheap human labor. Since non-lexical features are extracted from users' interactions with timeline posts, imitating normal users' behaviors will likely incur an remarkably higher cost. Even higher would be the cost to interfere with lexical features, since this would entail modifying or imitating normal users' writing style.

\bibliographystyle{abbrv}

\end{document}